\documentclass[twocolumn,trackchanges]{aastex701} %linenumbers
\usepackage{amsmath,amssymb,graphicx,xcolor}
\usepackage{float}
\usepackage{silence}
\begin{document}

\title{ALMA’s sharpest view of IRAS 08544$-$4431: \\Unveiling the dust distribution asymmetries in the circumbinary disk}

\author[orcid=0000-0002-7453-2945,sname='Mohorian']{Maksym Mohorian}
\affiliation{School of Mathematical and Physical Sciences, Macquarie University, Balaclava Road, Sydney, NSW 2109, Australia}
\affiliation{Astrophysics and Space Technologies Research Centre, Macquarie University, Balaclava Road, Sydney, NSW 2109, Australia}
\email[show]{maksym.mohorian@mq.edu.au}  

\author[orcid=0000-0001-8299-3402,gname=Devika,sname=Kamath]{Devika Kamath} 
%\altaffiliation{Las Campanas Observatory}
\affiliation{School of Mathematical and Physical Sciences, Macquarie University, Balaclava Road, Sydney, NSW 2109, Australia}
\affiliation{Astrophysics and Space Technologies Research Centre, Macquarie University, Balaclava Road, Sydney, NSW 2109, Australia}
\affiliation{INAF, Osservatorio Astronomico di Roma, Via Frascati 33, I-00077 Monte Porzio Catone, Italy}
\email{devika.kamath@mq.edu.au}

\author[orcid=0009-0004-2905-6515,sname=Elizabeth,gname=Cappellazzo]{Elizabeth Cappellazzo}
\affiliation{School of Mathematical and Physical Sciences, Macquarie University, Balaclava Road, Sydney, NSW 2109, Australia}
\affiliation{Astrophysics and Space Technologies Research Centre, Macquarie University, Balaclava Road, Sydney, NSW 2109, Australia}
\affiliation{ATNF, CSIRO, Space and Astronomy, PO Box 76, Epping, NSW 1710, Australia}
\email{elizabeth.cappellazzo@hdr.mq.edu.au}

\author[orcid=0000-0002-3780-4112,gname=Kateryna,sname=Andrych]{Kateryna Andrych}
\affiliation{School of Mathematical and Physical Sciences, Macquarie University, Balaclava Road, Sydney, NSW 2109, Australia}
\affiliation{Astrophysics and Space Technologies Research Centre, Macquarie University, Balaclava Road, Sydney, NSW 2109, Australia}
\email{kateryna.andrych@mq.edu.au}

\author[orcid=0000-0002-9491-393X,gname=Jacques,sname=Kluska]{Jacques Kluska}
\affiliation{Institute of Astronomy, KU Leuven, Celestijnenlaan 200D, 3001 Leuven, Belgium}
\email{jacques.kluska@protonmail.com}

\author[orcid=0000-0001-5158-9327,gname=Hans,sname='van Winckel']{Hans van Winckel}
\affiliation{Institute of Astronomy, KU Leuven, Celestijnenlaan 200D, 3001 Leuven, Belgium}
\email{hans.vanwinckel@kuleuven.be}

\author[orcid=0000-0002-4716-4235,gname=Daniel,sname=Price]{Daniel Price}
\affiliation{School of Physics and Astronomy, Monash University, Wellington Rd, Clayton VIC 3800, Australia}
\affiliation{Univ. Grenoble Alpes, CNRS, IPAG, 38000 Grenoble, France}
\email{daniel.price@monash.edu}

\author[orcid=0000-0001-5760-7557,gname=Toon,sname='De Prins']{Toon De Prins}
\affiliation{School of Mathematical and Physical Sciences, Macquarie University, Balaclava Road, Sydney, NSW 2109, Australia}
\affiliation{Astrophysics and Space Technologies Research Centre, Macquarie University, Balaclava Road, Sydney, NSW 2109, Australia}
\affiliation{Institute of Astronomy, KU Leuven, Celestijnenlaan 200D, 3001 Leuven, Belgium}
\email{toon.deprins@kuleuven.be}

\author[orcid=0000-0001-6872-4206,gname=Arancha,sname=Castro Carrizo]{Arancha Castro-Carrizo}
\affiliation{Institut de Radioastronomie Millim\'{e}trique, 300 rue de la Piscine, 38406 Saint-Martin-d'H\`{e}res, France}
\email{ccarrizo@iram.fr}

\author[orcid=0000-0003-1968-0117,gname=Javier,sname=Alcolea]{Javier Alcolea}
\affiliation{Observatorio Astronómico Nacional (OAN-IGN), Alfonso XII 3, 28014 Madrid, Spain}
\email{j.alcolea@oan.es}

\author[orcid=0000-0003-3713-8073,gname=Nicol\'{a}s,sname=Cuello]{Nicol\'{a}s Cuello}
\affiliation{Univ. Grenoble Alpes, CNRS, IPAG, 38000 Grenoble, France}
\email{nicolas.cuello@univ-grenoble-alpes.fr}

\author[orcid=0000-0003-4200-7852,gname=Matthias,sname=Fabry]{Matthias Fabry}
\affiliation{Department of Astrophysics and Planetary Science, Villanova University, 800 E. Lancaster Ave., Villanova, PA 19085, USA}
\email{matthias.fabry@villanova.edu}

\author[orcid=0000-0002-6586-4665,gname=Valentin,sname=Bujarrabal]{Valentin Bujarrabal}
\affiliation{Observatorio Astronómico Nacional (OAN-IGN), Apartado 112, 28803 Alcalá de Henares, Spain}
\email{v.bujarrabal@oan.es}

%% Use the \collaboration command to identify collaborations. This command
%% takes an optional argument that is either a number or the word "all"
%% which tells the compiler how many of the authors above the command to
%% show. For example "\collaboration[all]{(DELVE Collaboration)}" wil include
%% all the authors above this command.
%%
%% Mark off the abstract in the ``abstract'' environment. 
\begin{abstract}
The interactions between post-asymptotic giant branch (post-AGB) binaries and their circumbinary disks play a central role in shaping the evolution of these systems. Circumbinary disks around post-AGB binaries are stable, though relatively short-lived structures ($10^4-10^5$ years) and their physical properties and evolution remain poorly understood. In this study, we target IRAS\,08544$-$4431, one of the best-studied post-AGB binaries. This system has well-constrained stellar and orbital parameters, exhibits well-defined characteristic refractory-depleted photospheric chemistry, and hosts a stable, sub-Keplerian circumbinary disk extending from a well-resolved sublimation rim at approximately 8\,au to outer radii of approximately 1\,000\,au. We probe the disk layers close to the midplane using ALMA Band~7 continuum and $^{12}$CO $J=3-2$ observations of IRAS\,08544$-$4431 obtained with baselines up to 16.2\,km. These data provide the finest angular resolution yet achieved with ALMA for dust-continuum observations of post-AGB binaries with a (almost) face-on disk (15\,au; 10\,mas). The resulting map of dust continuum enabled resolving disk inner rim, a largely smooth radial profile, and a prominent azimuthal asymmetry at a radius of $\sim$45\,mas. Notably, the location of this dust asymmetry coincides with the forward-scattering peak reported in SPHERE/VLT observations of IRAS\,08544$-$4431. The $^{12}$CO emission is more extended than the dust continuum and confirms the disk rotation in this system. However, the global $^{12}$CO map shows indications of a spatial offset from the dust asymmetry, suggesting some degree of gas-dust decoupling. Together, these observations establish IRAS\,08544$-$4431 as a key laboratory for studying disk and dust evolution in a parameter space distinct from that of young stars with protoplanetary disks.
\end{abstract}
%% Keywords should appear after the \end{abstract} command. 
%% The AAS Journals now uses Unified Astronomy Thesaurus (UAT) concepts:
%% https://astrothesaurus.org
%% You will be asked to selected these concepts during the submission process
%% but this old "keyword" functionality is maintained in case authors want
%% to include these concepts in their preprints.
%%
%% You can use the \uat command to link your UAT concepts back its source.
% --- \uat{Cosmology}{343} --- \uat{High Energy astrophysics}{739} --- \uat{Interstellar medium}{847} --- \uat{Stellar astronomy}{1583} --- \uat{Solar physics}{1476}}

%% From the front matter, we move on to the body of the paper.
%% Sections are demarcated by \section and \subsection, respectively.
%% Observe the use of the LaTeX \label
%% command after the \subsection to give a symbolic KEY to the
%% subsection for cross-referencing in a \ref command.
%% You can use LaTeX's \ref and \label commands to keep track of
%% cross-references to sections, equations, tables, and figures.
%% That way, if you change the order of any elements, LaTeX will
%% automatically renumber them.

\section{Introduction}\label{sec:int}
The Atacama Large Millimeter/submillimeter Array (ALMA) significantly advanced the study of circumstellar environments during the final stages of stellar evolution \citep[e.g.,][]{olofsson2015HD101584, olofsson2019HD101584, agundez2017CarbonDust, nisini2018TTauJetsWindsAccretion, decin2015CWLeoSpiralsWind, decin2017AlAGB, decin2020StellarWinds}. Leveraging sub-arcsecond angular resolution \citep[down to $\sim$10\,mas in mm/sub-mm regime;][]{wootten2009ALMA}, ALMA has enabled detailed imaging of emission in both continuum and molecular bands, tracing circumstellar dust and gas, respectively. These capabilities have allowed investigations of the morphology and kinematics of circumstellar material around a wide variety of evolved objects, including asymptotic giant branch (AGB) stars \citep[e.g.,][]{agundez2017CarbonDust, decin2017AlAGB, khouri2016ALMACO5AGB, khouri2018MiraALMA}, red supergiants \citep[e.g.,][]{ogorman2015ALMARSGVYCMa, ogorman2017ALMARSGBetelgeuse, decin2024ALMARSG}, post-asymptotic giant branch (post-AGB) and post-red giant branch (post-RGB) binaries\footnote{For convenience, and given the strong similarities in observed stellar parameters between post-AGB and post-RGB binaries \citep[except for the luminosity range and nucleosynthetic history; see, e.g.,][]{vanwinckel2025Review}, we hereafter refer to both classes collectively as post-AGB binaries.} \citep[e.g.,][]{bujarrabal2013RedRectangle, bujarrabal2015KeplerianRotation, bujarrabal2018IRAS08, olofsson2015HD101584, olofsson2019HD101584, khouri2025ALMACOin6pAGB}, and planetary nebulae \citep{schmidt2016ALMAPN, santandergarcia2017ALMAPN}.

Post-AGB binaries represent a short-lived \citep[$10^4-10^5$ years;][]{millerbertolami2016DurationPostAGB} evolutionary stage of low- and intermediate-mass stars ($\sim0.8-8\,M_{\odot}$) after the star leaves the asymptotic giant branch but before it becomes a white dwarf ionizing its circumstellar environment \citep[observed as a planetary nebula; see, e.g.,][]{vanwinckel2003PostAGBReview}. In binary systems, the AGB evolution might be prematurely terminated by interactions between binary components: a post-AGB primary and a main-sequence companion \citep{oomen2018OrbitalProperties, vanwinckel2025Review}. These binary interactions play a key role in the formation of circumbinary disks surrounding the central binary \citep[CBDs; e.g.,][]{vanwinckel2003PostAGBReview, vanwinckel2009Binarity, kamath2016PostRGBDiscovery}.

The presence of CBDs in post-AGB systems was initially inferred from their spectral energy distributions (SEDs), which commonly display a broad infrared (IR) excess arising from hot dust located close to the central binary \citep[e.g.,][]{vanwinckel2003PostAGBReview, deruyter2006KeplerianDiscs, kamath2016PostRGBDiscovery}. The shape of the IR excess in the SEDs of post-AGB binaries enabled a phenomenological classification of CBDs into three types: full disks, transition disks, and faint disks \citep{kluska2022GalacticBinaries, corporaal2023FullDisc, corporaal2023TransitionDisc}. In full disks, the dust distribution extends outward starting from the sublimation rim \citep[SED excess starts in near-IR range;][]{kluska2018IRAS08}. In transition disks \citep{kenyon1995TransitionPPDs}, the dust distribution starts beyond the sublimation radius showing inner cavities of unknown origin potentially linked to the first- or second-generation planet formation \citep[SED excess starts in mid-IR range;][]{kluska2022GalacticBinaries, corporaal2023TransitionDisc}. In both full and transition disks, a high fraction of the stellar irradiation is re-emitted in the IR range (IR-to-stellar luminosity ratios of $L_{\rm IR}/L_\ast>0.1$), requiring a significant scale height in these disks \citep[H/r$>$0.1;][]{vanwinckel2007DiskScaleHeight, andrych2023DiscStructuresIRDIS}. Finally, in faint disks, the dust content is either low or settled in the midplane \citep[SED excess is significantly weaker, as evidenced by IR-to-stellar luminosity ratios of $L_{\rm IR}/L_\ast<0.1$;]{mohorian2025FaintDiscsDepletion}.

Over the past decade, high-resolution polarimetric scattered-light imaging with SPHERE/VLT and high-resolution imaging with the Very Large Telescope Interferometer (VLTI) revealed that full and transition disks exhibit complex morphologies, including rings, cavities, and inner gaps \citep[e.g.,][]{andrych2023DiscStructuresIRDIS, andrych2024IRAS08, andrych2025DiscStructuresZIMPOL, corporaal2023TransitionDisc}. Furthermore, molecular line observations with radio-interferometric facilities (e.g., ALMA; Northern Extended Millimeter Array, NOEMA; Submillimeter Array, SMA) have demonstrated that the circumstellar environments of post-AGB binaries consist of CBDs in stable (sub-)Keplerian rotation and low-velocity expanding outflows \citep[1-10 km s$^{-1}$; see, e.g.,][]{bujarrabal2015KeplerianRotation, bujarrabal2018IRAS08, gallardocava2021PostAGBOutflows, gallardocava2023thesis}. Finally, photometric and spectroscopic studies of post-AGB systems in the IR regime showed evidence of dust processing within the disks, including strong crystallization, up to $\sim$70\%, and grain growth, further supporting the interpretation of CBDs around post-AGB binaries as second-generation PPDs \citep{gielen2011FullerenesPAHs, gielen2011Silicates, scicluna2020GrainGrowth, bujarrabal2023RedRectangle}.

The CBD interacts with its central binary primarily through the re-accretion of disk matter \citep{vanwinckel2003PostAGBReview, vanwinckel2009Binarity}. The re-accreted matter is chemically altered by the processes occurring within the CBD, involving gas condensation into dust and subsequent gas-dust fractionation: elements with high condensation temperatures remain locked in the disk (e.g., Al, Fe, and Ti; refractory elements), whereas elements with low condensation temperatures are re-accreted onto the central binary \citep[e.g., Na, S, Zn; volatile elements; see][and references therein]{mohorian2025FaintDiscsDepletion}. For a primary star with a thinning convective envelope, re-accreted refractory-poor disk matter dominates the initial surface composition, leading to the commonly observed underabundance of refractory elements in the stellar photosphere \citep[photospheric chemical depletion; see, e.g.,][]{waters1992Depletion, mohorian2025TransitionDiscsDepletion, mohorian2025FaintDiscsDepletion}. For the secondary companion, re-accretion leads to the formation of a circumsecondary disk, which is a crucial prerequisite for launching the observed high-velocity bipolar outflows \citep[jets; see, e.g.,][]{bollen2022JetStructure, verhamme2024Jets, deprins2024JetFormation, alcolea2025BipolarJet}. These processes illustrate ongoing dynamical interactions between the CBD and the central binary and might lead to the observed disk asymmetries.

Within the currently known sample of 85 Galactic post-AGB binaries \citep{kluska2022GalacticBinaries}, the Red Rectangle is the first and so far the only system that has been observed in ALMA Band~7 continuum at the highest angular resolution currently achievable in the sub-mm regime \citep[$\sim20$\,mas;][]{bujarrabal2023RedRectangle, alcolea2025BipolarJet}. These observations enabled detailed investigation of the radial and vertical disk structure, revealing asymmetry between the eastern and western sides of the disk, a sharply truncated outer disk edge, and providing evidence for a molecular wind launched from the disk surface. The mass carried by this outflow appears to be small compared to that stored in the disk itself, with an estimated outflow-to-disk mass ratio of $\sim11\%$ \citep{gallardocava2023thesis}. However, since the Red Rectangle is viewed edge-on \citep{alcolea2025BipolarJet}, information on the azimuthal dust distribution, including reported asymmetries, remains limited for this target.

IRAS\,08544$-$4431 is another well-studied benchmark system among post-AGB binaries. This object has an almost face-on disk \citep[inclination of 19\,$\pm$\,2$^\circ$;]{hillen2016IRAS08, kluska2018IRAS08} and exhibits many characteristic observational signatures of a post-AGB binary \citep{maas2003IRAS08, deruyter2006KeplerianDiscs, vanwinckel2009Binarity}, including a strong IR excess in the SED \citep[$L_{\rm IR}/L_\ast$=0.29;][]{kluska2022GalacticBinaries}, orbital period of 501.1$\pm$1.0 d and eccentricity of 0.20$\pm$0.02 \citep{oomen2018OrbitalProperties}, outflow-to-disk mass ratio of $\sim10\%$ \citep{gallardocava2023thesis}, and photospheric depletion \citep{deruyter2006KeplerianDiscs}. Moreover, near-IR interferometric imaging of IRAS\,08544$-$4431 with VLTI enabled resolving the hot near-circular inner dust rim of the CBD and constraining the binary separation \citep[diameter of 14.15\,$\pm$\,0.10\,mas and position angle of 6\,$\pm$\,6$^\circ$ for the dust inner rim; angular separation of 0.81$\pm$0.05 mas for the central binary;][]{hillen2016IRAS08}. Furthermore, high-resolution polarimetric imaging of IRAS\,08544$-$4431 with SPHERE/VLT allowed to resolve the extended structure of the disk surface and to reveal significant forward scattering, suggesting that the northern part of the disk is closer to the observer. Additionally, ALMA observations of $^{12}$CO and $^{13}$CO emission (v=0 $J$=3-2) in IRAS\,08544$-$4431 enabled the estimation of a total stellar mass of the central binary (1.8\,$M_\odot$) and a total nebular mass from $\sim6\times10^{-3}\,M_{\odot}$ to $\sim2\times10^{-2}\,M_{\odot}$ \citep[for distances from 550\,pc to 1100\,pc\footnote{We note that the updated distance to IRAS\,08544--4431 based on the \textit{Gaia} DR3 parallaxes is 1575\,pc (see Section~\ref{ssec:resflx}).}; assuming X($^{12}$CO)=1.5$\times10^{-4}$ and X($^{13}$CO)=1.5$\times10^{-5}$]{bujarrabal2018IRAS08}. However, the spatial resolution of the available ALMA observations ($\sim0.1-0.2$\,arcsec) was insufficient to resolve any potential disk substructures in the dust continuum emission of IRAS\,08544$-$4431.

As disk–binary interactions are expected to shape the evolution, morphology, and chemistry of post-AGB systems, direct observational constraints at the relevant spatial scales are needed to probe the mechanisms of these interactions. In this study, we analyze high-resolution ALMA Band~7 observations that resolve the circumbinary disk at spatial scales comparable to the diameter of the disk sublimation rim. These data allow us to investigate the detailed structure of the disk and to search for signatures of asymmetries that may reflect ongoing disk–binary interaction. The structure of this paper is as follows. In Section~\ref{sec:obs}, we describe the observations and data reduction. In Section~\ref{sec:res}, we present the data analysis and results. In Section~\ref{sec:dsc}, we discuss the implications of the observed asymmetries and compare these features with those observed in other systems, including the Red Rectangle and young stars hosting protoplanetary disks (PPDs). In Section~\ref{sec:con}, we summarize the main conclusions of this study.

\section{Observations and Data Reduction}\label{sec:obs}
IRAS 08544$-$4431 was observed with ALMA in Band 7 in six observation blocks: three in 2015 (29 and 30 August; ID 2013.1.00338.S; PI Bujarrabal), two in 2016 (25 August and 3 September; ID 2013.1.00338.S; PI Bujarrabal), and one in 2021 (5 September; ID 2019.1.00919.S; PI Kluska). The array configuration provided baselines ranging from $\sim$200 m to $\sim$16 km. The average integration time was 3.15 hours in the 2015-2016 blocks and 49 minutes in the 2021 block. For the 2015-2016 observations, the correlator configurations were detailed in \citet{bujarrabal2018IRAS08}. For the 2021 observation, the correlator was configured with 1920 channels in each of the four spectral windows centered at 347, 349, 357, and 359 GHz, respectively (channel width of 976.562 kHz, $\sim$0.9\,km s$^{-1}$).

We reduced and calibrated the 2021 data using the standard ALMA pipeline and Common Astronomy Software Applications for Radio Astronomy \citep[CASA, version 6.6.1;][]{bean2022CASA}, employing pipeline reduction with manual flagging of poor visibilities. The quasars J0538-4405, J0849-3541, and J0922-3959 were used as the bandpass and flux calibrator (with integrated flux of 1.561 Jy), water vapour radiometer calibrator (with integrated flux of 0.276 Jy), and phase calibrator (with integrated flux of 0.810 Jy), respectively. We note that quasar J0538-4405 was used as the flux calibrator in all observing blocks from 2015 to 2021 (0.886/1.356/1.561 Jy in 2015/2016/2021, respectively). We verified that the used flux reference of 1.56 Jy was reliable for J0538-4405 with an error estimated at $<$10\% (based on a large number of consistent ALMA measurements in weeks surrounding the 2021 observation). The final continuum maps of IRAS\,08544$-$4431 were produced by combining the line-free channels across all spectral windows and by using pixel (cell) size of 2 mas with the \texttt{tclean} task with multiscale recipe \citep{cornwell2008Multiscale}, Briggs weighting \citep{briggs1995Weighting}, and a robust parameter of 0.0. Continuum subtraction was performed in the visibility domain using line-free channels. The continuum-subtracted data cube was used to produce the $^{12}$CO line map. The final $^{12}$CO line maps (including the moment 0 map) of IRAS\,08544$-$4431 were produced using the \texttt{tclean} task with multiscale recipe \citep{cornwell2008Multiscale}, natural weighting, and binned pixel size of 10 mas.

For the 2015-2016 datasets, our final maps of IRAS\,08544$-$4431 are similar to those presented in \citet{bujarrabal2018IRAS08}, where an angular resolution of 120$\times$140 mas precludes from resolving the inner regions of the CBD. For the 2021 dataset, we achieved an angular resolution of 9.7$\times$11.5 mas and an RMS noise of 0.06 mJy beam$^{-1}$ in the continuum maps and 1.5 mJy beam$^{-1}$ per 1 km s$^{-1}$ channel in the spectral line cubes. In Appendix~\ref{app:rob}, we provide the ALMA maps of continuum and $^{12}$CO line obtained with natural, Briggs (robust of 0.0), and uniform weightings for the combined (2015, 2016, and 2021) dataset. The 2021 dataset improves the angular resolution of available continuum data for IRAS\,08544$-$4431 by a factor of $\sim$10, allowing us to directly resolve the sublimation rim in the dust continuum (which has a diameter of 14.15$\pm$0.10\,mas; see Section~\ref{sec:int}). The spectral resolution of the 2021 observation is significantly lower than that of the 2015-2016 observations (0.8 km s$^{-1}$ and 0.1 km s$^{-1}$, respectively), limiting the analysis of the gas evolution in the CBD of IRAS\,08544$-$4431. In this study, we present the continuum and $^{12}$CO line map for the 2021 observation, as well as the continuum map for the combined 2015-2021 dataset.

\begin{figure}[htb]
    \centering
    \includegraphics[trim={1cm 1cm 0.8cm 1cm}, width=\columnwidth]{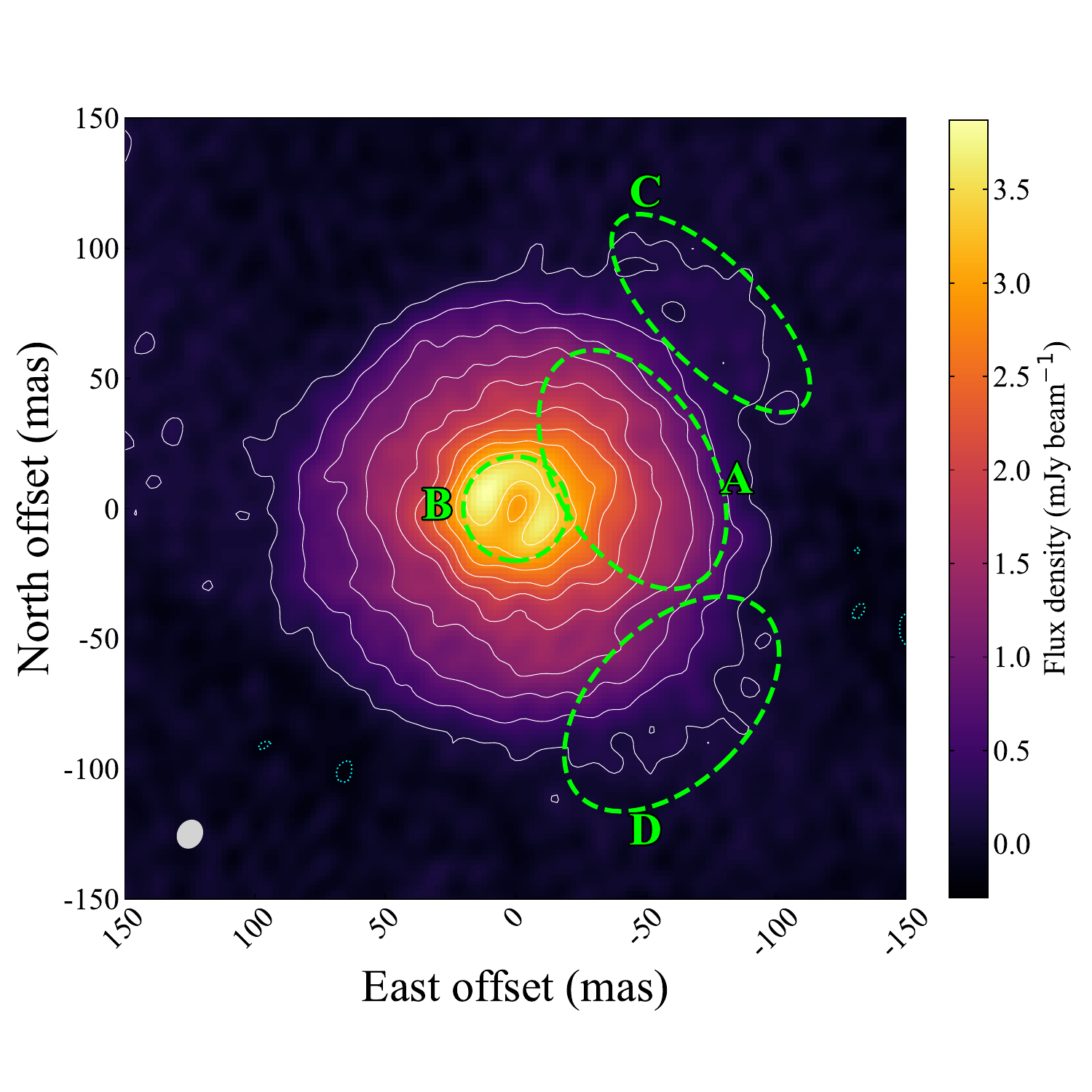}
    \caption{The 2021 continuum map of IRAS\,08544$-$4431. The beam size and orientation are shown in the bottom left corner (9.7$\times$11.5 mas, $PA$=-27.77$^\circ$; minimum and maximum flux densities are -0.28 and 3.87 mJy beam$^{-1}$, respectively). The lowest-flux-density contour is 0.18 mJy beam$^{-1}$ (3$\times$RMS), the highest-flux-density contour is 3.42 mJy beam$^{-1}$ (57$\times$RMS), and the step is 0.36 mJy beam$^{-1}$ (6$\times$RMS); negative contours (-0.18 mJy beam$^{-1}$; -3$\times$RMS) are indicated by dotted cyan lines. The non-axisymmetric features are annotated with dashed lime lines and lime letters (for more details, see Section~\ref{sec:res}).}\label{fig:cntmap}
\end{figure}
\begin{figure}[htb]
    \centering
    \includegraphics[trim={1cm 0.8cm 0.8cm 1cm}, width=\columnwidth]{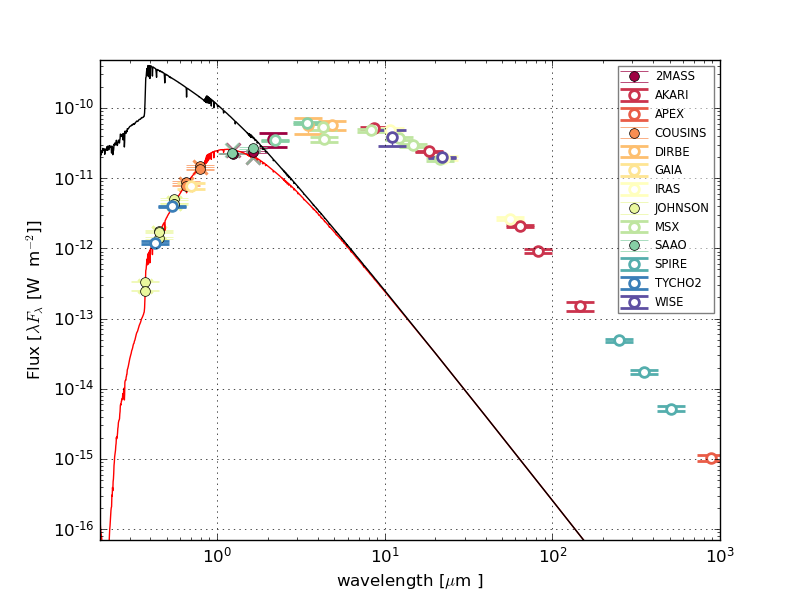}
    \caption{Spectral energy distribution of IRAS\,08544$-$4431. The legend shows all catalogs, the photometric data points from which were adopted in this study. The red solid line marks the fitted Kurucz model atmosphere, the black solid line marks the de-reddened model atmosphere. The interpolation of the long-wavelength ($\lambda$\,$>$\,50\,$\mu$m) SED tail provides the spectral index of -1.90$\pm$0.05 and the expected total flux in ALMA Band~7 of 327.4 mJy beam$^{-1}$ (for more details, see Section~\ref{ssec:resflx}).}\label{fig:fitsed}
\end{figure}

\section{Data Analysis and Results}\label{sec:res}
In this section, we present the results of our analysis for the 2021 ALMA data of IRAS\,08544$-$4431. In Section~\ref{ssec:resflx}, we report the calculated values for total continuum flux and dust mass. In Section~\ref{ssec:resasy}, we describe the detected asymmetries in the dust continuum. In Section~\ref{ssec:resgas}, we present the gas distribution traced by the $^{12}$CO line map.

\subsection{Total Flux and Dust Mass}\label{ssec:resflx}
The 0.87-mm ALMA continuum of IRAS\,08544$-$4431 in the 2021 dataset enables resolving the disk structure and measuring the total mass of mm/sub-mm-sized dust grains. In Fig.~\ref{fig:cntmap}, we show the 2021 continuum map of IRAS\,08544$-$4431 and four non-axisymmetric features (see Section~\ref{ssec:resasy}). From this continuum map, we measure a total flux of 271.4\,$\pm$\,1.5\,mJy, integrated over emission above 0.3 mJy beam$^{-1}$ (5$\times$RMS). For photometric data points with $\lambda$$>$50$\mu$m, the SED fit (see Fig.~\ref{fig:fitsed}) provides the spectral index of --1.90$\pm$0.05, predicting the total flux of 327.4$\pm$\,8.2\,mJy within the observed frequency range. For the 2015/2016 ALMA observations, \citet{bujarrabal2018IRAS08} reported a total flux of $\sim$320\,$\pm$\,1\,mJy and an emission size of $\sim$0.1\,arcsec, consistent with the more compact baseline coverage of 2015/2016 data. The lower recovered flux in the 2021 observations therefore suggests that the missing fraction of the continuum flux ($\sim$17\%) arises from regions larger than $\sim$0.2\,arcsec and is overresolved due to the extent of the baseline configuration (see Section~\ref{sec:obs}).

The calculated total continuum flux of 271.4\,$\pm$\,1.5\,mJy enables estimation of the dust mass with a detected emission at radii up to $\sim$100 mas using the flux–mass relation for an optically thin disk\footnote{We note that the brightness temperature of the region with continuum fluxes above 5$\times$RMS in Fig.\ref{fig:cntmap} ranges from $\sim$30\,K to $\sim$400\,K, implying that the CBD of IRAS\,08544--4431 is optically thick for the 0.87-mm emission (further corroborated by the spectral index of $\sim$2). In this case, Equation~\ref{eq:mss} provides a lower limit on the dust mass.}, given as
\begin{equation}\label{eq:mss}
M_{\rm dust, thin} = \dfrac{F_\nu\times d^2}{\kappa_\nu\times B_\nu(T_{\rm dust})} < M_{\rm dust, thick},
\end{equation}
where $F_\nu$ is the integrated ALMA Band~7 flux, $d=1.575$\,kpc is the adopted distance to IRAS\,08544$-$4431 based on the \textit{Gaia} DR3 parallaxes \citep{andrych2024IRAS08,menon2024SEDfitting}, $\kappa_\nu$ is the dust opacity, and $B_\nu(T_{\rm dust})$ is the Planck function evaluated for the dust temperature $T_{\rm dust}$. As the observed dust continuum has different temperatures across the disk, the Planck function in Eq.~\ref{eq:mss} should be calculated for the mass-averaged $T_{\rm dust}$. However, the mass distribution of dust in the CBDs of post-AGB binaries is poorly constrained, so instead we assume two following marginal cases for $T_{\rm dust}$ to obtain the limits on the dust mass: $T_{\rm dust}$=1\,300\,K \citep[sublimation rim;][]{hillen2016IRAS08, kluska2019PostAGBDiscs, corporaal2023FullDisc} and $T_{\rm dust}$=50\,K \citep[derived from the disk model of][at radii of $\sim$100\,mas]{bujarrabal2018IRAS08}. We adopt dust opacity per unit dust mass $\kappa_\nu \approx 3.5$\,cm$^{2}$\,g$^{-1}$, appropriate for sub-mm-sized dust grains in PPDs observed with ALMA Band~7 \citep{birnstiel2018DustOpacityDSHARP, viscardi2025DustOpacity}, as CBDs of post-AGB binaries were reported to also host such large grains \citep{scicluna2020GrainGrowth}.

From Equation~\ref{eq:mss}, we infer an observed dust mass in IRAS\,08544$-$4431 in the range from $1.9\times10^{-4}$ to $5.7\times10^{-3}\,M_\odot$ (for the observed total flux of 271.4 mJy) and the total dust mass to have a lower limit in the range from $2.3\times10^{-4}$ to $6.9\times10^{-3}\,M_\odot$ (for the total SED-extrapolated flux of 327.4 mJy). Recently, the dust mass was also calculated for the Red Rectangle \citep[$\approx5\times10^{-5}-10^{-4}\,M_\odot$;][]{bujarrabal2023RedRectangle, alcolea2025BipolarJet}. Using Equation~\ref{eq:mss} and assuming optical thinness, we derive the dust mass limits of $1.0\times10^{-4}-3.1\times10^{-3}\,M_\odot$ for the Red Rectangle, which mostly overlaps with the limits obtained for IRAS\,08544$-$4431 in this study. This highlights that substantial diversity among post-AGB binary systems in stellar and binary observational parameters might not be as prominent in dust reservoirs. We note that our inferred dust mass for the CBD of the Red Rectangle is $\sim$8 times higher than those presented in \citet{bujarrabal2023RedRectangle, alcolea2025BipolarJet}. This discrepancy is significantly affected by the adopted distance used in the calculations of gas mass \citep[1.1\,kpc;][]{bujarrabal2023RedRectangle} and dust mass (1.575\,kpc; this study). A detailed re-analysis of the disk model and corresponding gas mass for the updated distance is beyond the scope of this study.

Furthermore, assuming a total gas mass of $M_{\rm gas}\sim2\times10^{-2}\,M_\odot$ for IRAS\,08544$-$4431 (see Section~\ref{sec:int}), the derived dust mass implies a gas-to-dust mass ratio $\Delta_{\rm gas/dust}$ in the range from $\sim$3 to $\sim$100, which is mostly below the commonly adopted mass ratio for circumbinary disks \citep[$\sim$100;][]{corporaal2023FullDisc}. Accounting for the distance discrepancy in gas mass and dust mass calculations increases the limits of gas-to-dust mass ratio $\Delta_{\rm gas/dust}$ by a factor of $\sim$2.25 to the range between $\sim$8 and $\sim$225). We also note that since optically thick dust emission may conceal additional dust mass, these gas-to-dust ratios should be regarded as upper limits. Hence, the derived mass values for both circumstellar gas and dust in IRAS\,08544$-$4431 should be interpreted with caution and remain subject to systematic uncertainties due to poorly constrained fractions of CO isotopologues to H$_2$ and dust optical thickness \citep{bujarrabal2015KeplerianRotation, scicluna2020GrainGrowth, gallardocava2023thesis}.

\subsection{Dust Continuum Asymmetries}\label{ssec:resasy}
\begin{figure*}[tb]
    \centering
    \includegraphics[width=\textwidth]{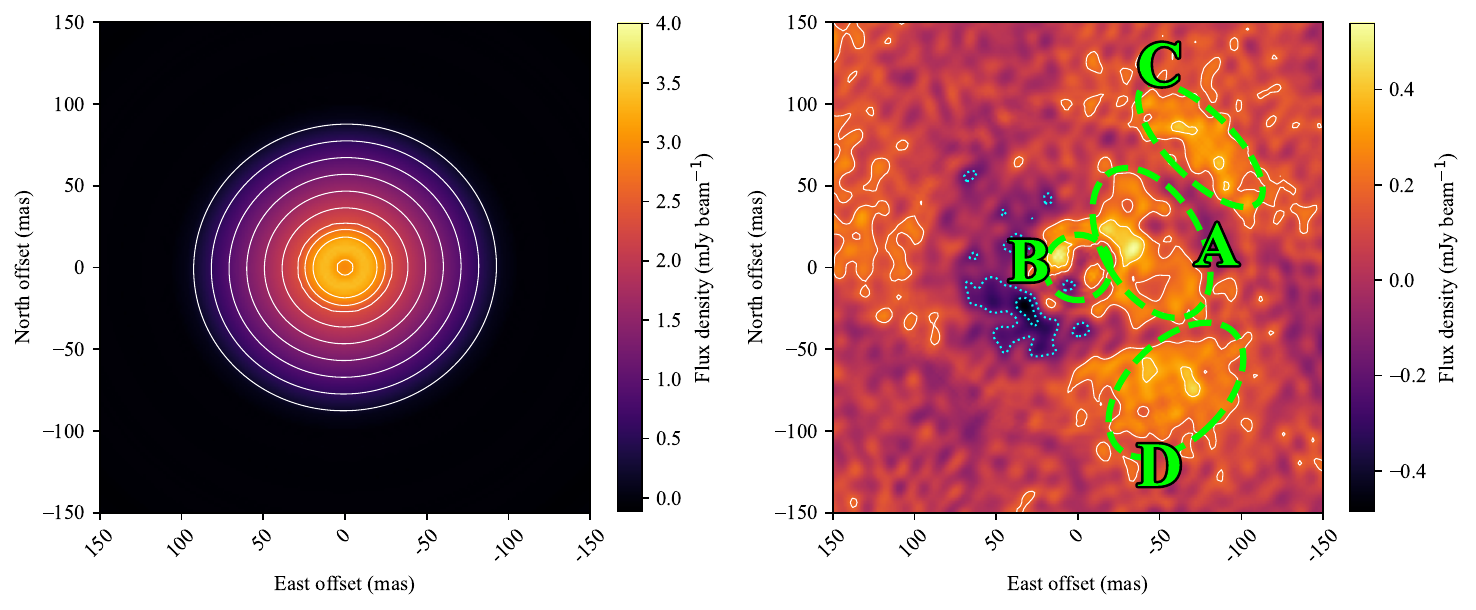}
    \caption{Azimuthally-averaged 16$^{th}$-percentile inclined disk profile of circumstellar emission in IRAS\,08544$-$4431 (\textit{left panel}) and subtracted map (residual map; \textit{right panel}) showing pronounced asymmetries of dust distribution (for more details, see Section~\ref{sec:res}). In the left panel, the lowest-flux-density contour is 0.18 mJy beam$^{-1}$ (3$\times$RMS), the highest-flux-density contour is 3.42 mJy beam$^{-1}$ (57$\times$RMS), and the step is 0.36 mJy beam$^{-1}$ (6$\times$RMS). In the right panel, the contours are $\pm$0.18 mJy beam$^{-1}$ ($\pm$3$\times$RMS) and $\pm$0.36 mJy beam$^{-1}$ ($\pm$6$\times$RMS); negative contours are indicated by dotted cyan lines. The non-axisymmetric features are annotated with dashed lime lines and lime letters, including the dominant non-axisymmetric feature of the dust distribution (structure A) with a flux density up to 20-25\% above the smooth disk profile, which corresponds to $>$6$\times$RMS (see Section~\ref{ssec:dscirs}).}\label{fig:radres}
\end{figure*}

To explore the symmetry of the observed ALMA dust continuum, we subtract an azimuthally-averaged 16$^{th}$-percentile inclined \citep[$i$=20$^\circ$, $PA$=6$^\circ$;][]{hillen2016IRAS08, bujarrabal2018IRAS08} disk profile from the observed dust continuum (see Fig.~\ref{fig:radres}, left panel). In the resulting residuals (see Fig.~\ref{fig:radres}, right panel), we detect structure A -- a pronounced enhancement (20-25\% above the averaged disk profile corresponding to $>$6$\times$RMS) extending from the northwest to the southwest at a distance of $\sim$70\,au ($\sim$45\,mas; see Section~\ref{ssec:dscirs}). This overdensity spans an azimuthal extent of $\sim$80\,au ($\sim$50\,mas) and is accompanied by a gap-like feature on the opposite (southeastern) side. In Section~\ref{ssec:dscirs}, we discuss the possible origin and significance of this asymmetry. We note that the almost face-on orientation of IRAS\,08544$-$4431 minimizes inclination-induced irregularities of the optical depth, which are therefore unlikely to account for the localized, azimuthally asymmetric overdensities observed in this study.

In addition to the dominant overdensity (structure A), the residual map of IRAS\,08544$-$4431 reveals larger but fainter asymmetric features, including localized asymmetric excesses at $r\approx$16-24\,au (10-15\,mas; structure B) coinciding with the sublimation rim resolved by near-infrared interferometry \citep{hillen2016IRAS08}. Weak residuals at $r\approx85$-$120$\,mas might trace the outer disk structure (structures C and D), possibly dynamically linked to the dominant overdensity or central binary through resonant interactions. The presence of perturbations over multiple radial scales suggests that the disk hosts a hierarchy of pressure variations, rather than a single isolated feature.

\subsection{Gas Distribution}\label{ssec:resgas}
\begin{figure*}[tb]
    \centering
    \includegraphics[trim={1cm 0.2cm 0.8cm 1cm}, width=\textwidth]{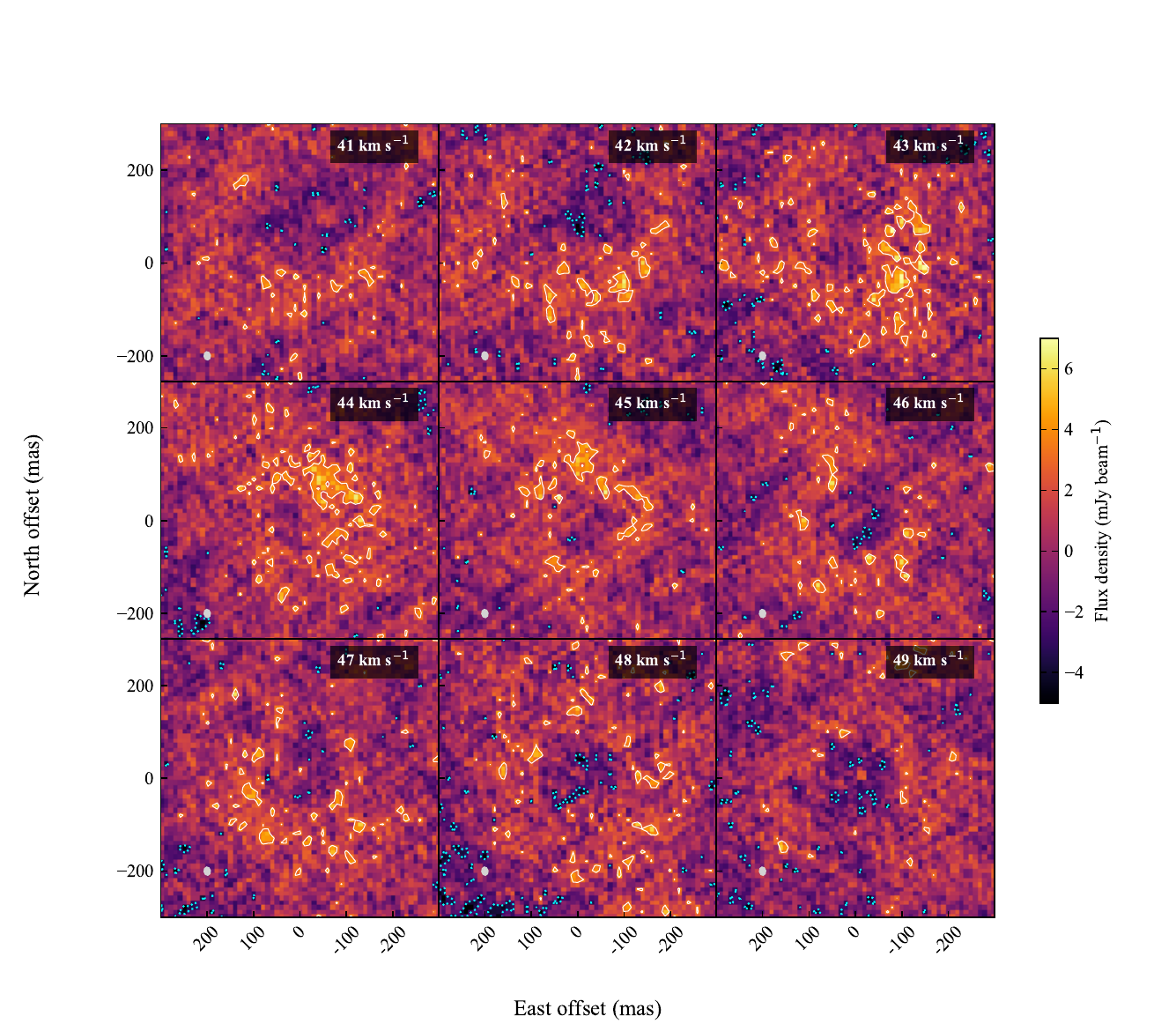}
    \caption{$^{12}$CO channel maps of IRAS\,08544$-$4431 from the 2021 data set. The LSR velocities are shown in the top right corners. The beam size and orientation are shown in the bottom left corner (16.4$\times$20.8 mas, $PA$=8.4$^\circ$). The contours mark the regions with flux densities above 3 mJy beam$^{-1}$ (2$\times$RMS; see Section~\ref{sec:obs}); negative contours (-3 mJy beam$^{-1}$; -2$\times$RMS) are indicated by dotted cyan lines. To increase S/N ratio of these maps, we used natural weighting and binned pixel size of 10 mas (for more details, see Section~\ref{sec:res}).}\label{fig:linmap}
\end{figure*}

\begin{figure}[tb]
    \centering
    \includegraphics[trim={0 0.25cm 0 0.25cm},width=\columnwidth]{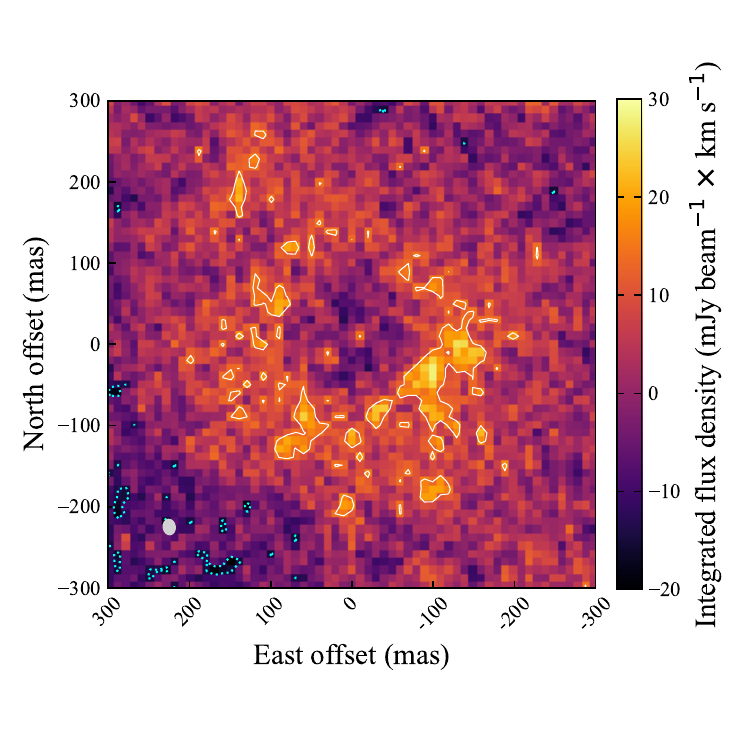}
    \caption{Global (moment 0) $^{12}$CO map of IRAS\,08544$-$4431 (integrated over 9 km s$^{-1}$) shows that $^{12}$CO emission is detected at radii of $\sim$80-100\,mas, which is significantly more extended than the continuum emission (see Fig.~\ref{fig:cntmap}). The beam size and orientation are shown in the bottom left corner (16.4$\times$20.8 mas, $PA$=8.4$^\circ$). The contours mark the regions with flux densities above 13.5 mJy beam$^{-1}$ km s$^{-1}$ (9$\times$RMS$\times$km s$^{-1}$); negative contours at --3 mJy beam$^{-1}$ km s$^{-1}$ (--2$\times$RMS$\times$km s$^{-1}$) are indicated by dotted cyan lines. To increase S/N ratio of this map, we used natural weighting and binned pixel size of 10 mas (for more details, see Section~\ref{sec:res}).}\label{fig:glbmap} %The peak of CO emission in the southwest direction deviates from the location of the continuum peak (northwest direction; see Section~\ref{sec:res}). The field orientation is shown in the bottom right corner
\end{figure}

The $^{12}$CO data cube from the 2021 dataset of IRAS\,08544$-$4431 shows larger radial extent of the $^{12}$CO gas, confirming the dust settling in the CBD of IRAS\,08544$-$4431 \citep[as shown in][]{bujarrabal2018IRAS08}. In Fig.~\ref{fig:linmap}, we show the $^{12}$CO line profiles with systemic velocity of $v_{\rm sys}$=45\,$\pm$\,1\,km s$^{-1}$ (local standard of rest), consistent with previous studies. The velocity structure of the disk's western side is consistent with previously reported disk rotation, although the low surface brightness ($<7.5$\,mJy beam$^{-1}$; $<5$\,RMS) and (almost) face-on orientation of the disk preclude a robust position-velocity analysis. In Fig.~\ref{fig:glbmap}, we present the global (moment~0) $^{12}$CO emission map, which shows that most of the detected CO flux is concentrated toward the southwestern side of the disk. The total $^{12}$CO flux detected in 2021 ALMA observation is 192.5\,$\pm$\,5.2\,mJy\,km\,s$^{-1}$ (integrated over emission above the 3$\times$RMS level). The southwestern gas concentration in the integrated $^{12}$CO map of IRAS\,08544$-$4431 lags behind the northwestern dust overdensity (structure A) given the rotation profile of this CBD \citep[see Fig.~\ref{fig:linmap} in this study and Fig. 1 in][]{bujarrabal2018IRAS08}, which suggests a possible spatial decoupling between the dust and gas components. In Section~\ref{ssec:dscirs}, we discuss this potential decoupling in more detail.


\section{Resolved Dust and Gas Distribution in the CBD of IRAS 08544--4431}\label{sec:dsc}
In this section, we examine the resolved non-axisymmetric features of the dust distribution (dust asymmetries) in the broader context of disk structure and evolution. In Section~\ref{ssec:dscirs}, we analyze the combined ALMA dataset of IRAS\,08544$-$4431 (Cycle 2 and Cycle 8) and discuss the significance of the detected overdensities in the dust distribution within the CBD with the dominant overdensity =(structure A) reaching flux density of 20-25\% above the smooth axisymmetric averaged disk profile (which corresponds to $>$6$\times$RMS level; see Section~\ref{ssec:resasy}). In Section~\ref{ssec:dscbin}, we compare our results with observed continuum observations of other post-AGB binaries. In Section~\ref{ssec:dscppd}, we draw parallels between the observed dust asymmetry in the CBD of IRAS\,08544$-$4431 and reported dust asymmetries in the PPDs of young stars.

\subsection{Dust Distribution Asymmetries in the CBD of IRAS 08544--4431}\label{ssec:dscirs}
\begin{figure*}[tbp]
    \centering
    \includegraphics[trim={1cm 1cm 0.8cm 1cm}, width=\textwidth]{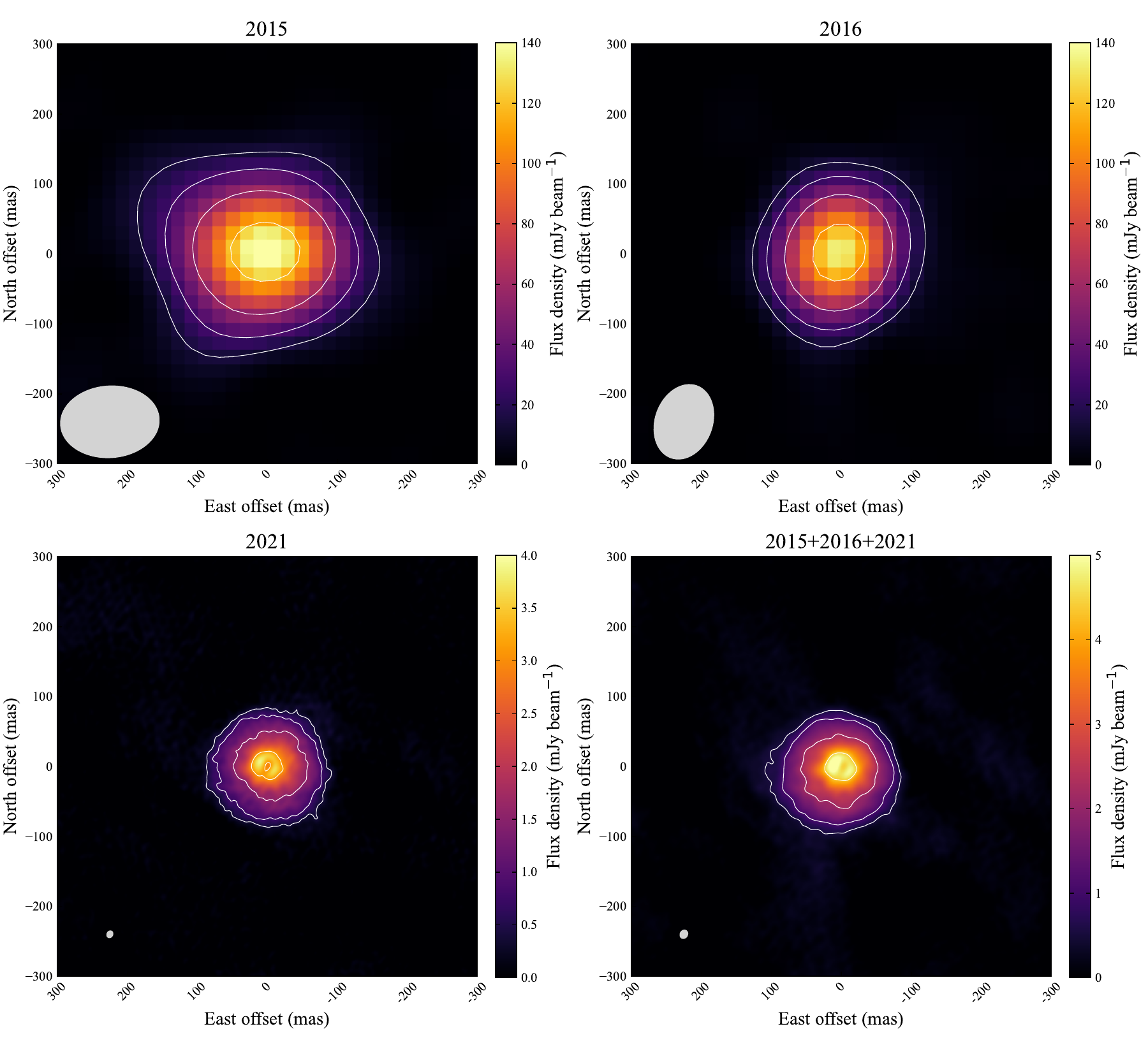}
    \caption{Yearly continuum maps of IRAS\,08544$-$4431: Cycle 2 (\textit{top panels}), Cycle 8 (\textit{bottom left panel}), and the combined data set (\textit{bottom right panel}). The yearly maps were derived using Briggs weighting with robust=0, and the combined map was derived using natural weighting. The beam size and orientation are shown in the bottom left corner. The contours mark the regions with flux densities of 10\%, 20\%, 40\%, and 80\% of the maximum flux density (with peak flux densities of 146.3, 133.4, 3.87, and 5.46 mJy beam$^{-1}$ for 2015, 2016, 2021, and combined dataset, respectively). For more details, see Section~\ref{ssec:dscirs}.}\label{fig:yrmaps}
\end{figure*}

Asymmetries in dust distribution had already been established for the CBD of IRAS\,08544$-$4431 both in the disk surface layers by SPHERE/VLT polarimetric scattered-light observations \citep[][see Section~\ref{sec:int} and Appendix~\ref{app:sph}]{andrych2024IRAS08} and in the inner rim by PIONIER/VLTI near-IR interferometric observations \citep{hillen2016IRAS08, kluska2018IRAS08}. In this study, the long-baseline 2021 ALMA data now extend that picture by resolving the distribution of dust grains in the disk layers closer to the midplane. These data achieve a resolution of $\sim$10\,mas, representing an order-of-magnitude improvement over the 2015–2016 ALMA observations (see Fig.~\ref{fig:yrmaps}). While the earlier ALMA epochs traced the global radial dust reservoir and were consistent with smooth radial distributions inferred from radiative transfer modeling of \citet{bujarrabal2018IRAS08}, the new data also reveal clear azimuthal departures from axisymmetry.

The dominant asymmetric feature in the dust continuum of IRAS\,08544$-$4431 (structure A) is located at $r\approx40$\,mas to the northwest, with an azimuthal extent of $\sim$50\,mas. This feature might be associated with a localized pressure maximum capable of trapping a small population of mm-sized grains (supported by the SED spectral index of --1.90$\pm$0.05; see Section~\ref{ssec:resflx}). In such a pressure maximum, dust particle distribution from moderately inertial species (represented by Stokes numbers St=$\tau_p/\tau_f\sim$0.1, where $\tau_p$ is the particle relaxation time and $\tau_f$ is the characteristic time scale of the flow) to significantly inertial pebbles (represented by St$\sim$1) might accumulate azimuthally, thereby decreasing the local gas-to-dust ratio, as commonly reported in PPDs \citep{vandermarel2013DustTrapTransitionDisc, cassassus2015DustTraps, dullemond2018DustTrapping, rosotti2020DustTrappingPPDs, lee2022StokesRingsinPPDs} and associated with disk-binary interaction \citep{ragusa2017DustTraps, ragusa2020DustTraps, price2018DustTraps}. The presence of additional weaker non-axisymmetric features at larger distances from the central binary (structures B, C, and D; see Section~\ref{ssec:resasy}) further suggests that the disk might host multiple pressure perturbations. We note that departures from the adopted axisymmetric model might also be contributed by variations in temperature or optical depth, geometric effects of disk eccentricity \citep[less likely due to near-circularity of the inner rim;][]{kluska2018IRAS08}, and unresolved substructure. To distinguish between these interpretations, new sensitive ALMA observations of IRAS\,08544$-$4431 are required.

Furthermore, the dominant overdensity in the 0.87-mm ALMA dust continuum of IRAS\,08544$-$4431 (structure A) is positionally aligned with the scattering-intensity asymmetry previously observed in the disk surface with SPHERE/VLT in April 2018 \citep{andrych2024IRAS08}. This alignment might indicate that the two observations trace the same underlying non-axisymmetric dust structure, extending from the disk surface to the continuum-emitting layers closer to the midplane. An alternative interpretation is that the apparent alignment is coincidental. Since the SPHERE/VLT asymmetry is expected to be dominated by forward scattering from the inclined disk surface, the position angle of this asymmetry should remain fixed. By contrast, if the ALMA continuum asymmetry traces a physical dust concentration, it should orbit together with the local disk flow. At the radius of the continuum overdensity (structure A), $\sim$45\,mas (Fig.~\ref{fig:cntmap}), and assuming a central binary mass of 1.8\,$M_\odot$ and Keplerian rotation of the disk out to $\sim$250\,mas \citep{bujarrabal2018IRAS08}, the corresponding Keplerian orbital period is of order $\sim$400\,yr.

The origin of such a dust concentration remains uncertain. One possibility is dust trapping in a local gas-pressure maximum or other non-axisymmetric structure embedded within the Keplerian disk \citep{bai2010MidplaneDynamicsPPDs}. These structures can arise through mechanisms such as Rossby wave instability \citep[see][and references therein]{rosotti2020DustTrappingPPDs, wolfer2025exoALMAVortex} or binary-induced perturbations \citep[most prominent for disks with an eccentric inner rim at comparable radii; see, e.g.,][and references therein]{calcino2019PPDEccentricOverdensity, ragusa2021PPDBinaryInducedOverdensity, cuello2026BinaryInducedOverdensity}. However, the present data does not enable establishing which mechanism is uniquely favoured in driving the observed perturbations. Instead, this question requires dedicated radiative-transfer modelling of the disk parameters in IRAS\,08544$-$4431. Future high-angular-resolution ALMA observations could test these scenarios by measuring whether the continuum overdensity changes position angle over time \citep[as previously shown for LkH$\alpha$ 101 in][]{tuthill2002LkHa101}. For a Keplerian period of $\sim$400\,yr, the expected angular motion is modest, of order $\sim$10 degrees over a decade, so such a test would require careful multi-epoch astrometric comparison to distinguish between a fixed position angle and a measurable rotation.

Finally, the western side of $^{12}$CO $J = 3-2$ line map from the 2021 ALMA observation (see Fig.~\ref{fig:linmap}) is indicative of the global rotation of the disk previously reported by \citet{bujarrabal2018IRAS08} and demonstrates that the $^{12}$CO emission is undetected for radii closer than $\sim$80\,mas. However, a detailed position–velocity analysis of the 2021 dataset is precluded by the limited sensitivity ($\sim$0.7\,mJy beam$^{-1}$ for a spectral line 10 km s$^{-1}$ wide) and by flux loss due to large-scale emission ($\gtrsim$0.2 arcsec; see Section~\ref{sec:res}). The moment 0 $^{12}$CO map (see Fig.~\ref{fig:glbmap}) shows that the bulk of the CO flux detected above the 9$\times$RMS noise level per channel is concentrated at larger radii ($\sim$80\,mas) toward the southwest direction (indication of an offset from the northwestern structure A with a resolved inner rim detected as structure B at $\sim$15\,mas). Further ALMA monitoring of IRAS\,08544$-$4431 will also help robustly determine the behavior of the observed gas kinematics in relation to the dominant dust asymmetry detected in the ALMA continuum.

\subsection{Comparison of Continuum Structures in CBDs of Post-AGB Binaries}\label{ssec:dscbin}
High-resolution imaging observations and spectral-index analyses established that CBDs of post-AGB binaries commonly contain large grains, with typical sizes of 0.1-1 mm \citep{gielen2011FullerenesPAHs, gielen2011Silicates, scicluna2020GrainGrowth, andrych2025DiscStructuresZIMPOL}. In addition, ALMA continuum and $^{12}$CO line maps revealed optically thick circumstellar environments with radial extents reaching beyond $\sim$500\,au around post-AGB binaries, including IRAS\,08544$-$4431 \citep{sanchezcontreras2022ALMAPPNDisc, gallardocava2023thesis, bujarrabal2023RedRectangle}. These studies primarily constrained radial structure and grain growth, whereas azimuthal structure on $\lesssim$200\,au scales has remained largely inaccessible, stemming from the typically large distances of post-AGB binaries \citep[$\gtrsim 10^3$ pc;][]{gallardocava2023thesis, vanwinckel2025Review}.

 % The field orientation is shown in the bottom right corner.

CBDs around post-AGB binaries are also systematically accompanied by extended gas outflows traced in $^{12}$CO lines, with widely varying outflow-to-disk mass ratios, ranging from 5\% to 75\% \citep{olofsson2015HD101584, olofsson2019HD101584, gallardocava2023thesis}. This diversity suggests that angular momentum and mass are redistributed in post-AGB systems with highly variable efficiency. Therefore, an important unresolved question is whether the emergence of such prominent pressure-supported dust structures measurably influences the efficiency of disk-driven gas outflows. If this correlation exists, it would provide a valuable empirical constraint on the physical mechanisms governing disk-binary interaction and angular momentum redistribution in evolved binary systems.

In this study, the detection of a dust asymmetry at $\sim$40-mas separation (i.e. dominant dust asymmetry) in the ALMA continuum map of IRAS\,08544$-$4431 (see Section~\ref{ssec:dscirs}) shows that CBDs of post-AGB binaries, beyond the sublimation rim, might show significant (up to 25\% in flux ratio corresponding to $>$6$\times$RMS level) asymmetries on the otherwise radially smooth profiles. Comparable dust-distribution asymmetries might also be present in the 0.87-mm continuum map of the Red Rectangle, the only other post-AGB binary studied at similar scales in the sub-mm regime \citep[angular resolution of $\sim$20 mas;][]{alcolea2025BipolarJet}, although its almost edge-on orientation precludes a robust assessment of azimuthal symmetry. Nevertheless, the ALMA continuum emission of the Red Rectangle is also dominated by large grains, and its spatially resolved structure shows that advanced grain processing can persist over the relatively short evolutionary timescales of the post-AGB phase ($\sim 10^4-10^5$ yr; see Section~\ref{sec:int}).

Finally, we note that mm- and sub-mm-sized dust grains are inferred to be ubiquitous in CBDs of post-AGB binaries from far-IR/sub-mm spectral indices \citep{scicluna2020GrainGrowth}, indicating that efficient grain growth and (quasi-)stable fractionation of gas and dust occur in the CBDs of the entire population of post-AGB binaries \citep{oomen2020DiscBinaryInteractions}. Furthermore, the observed photospheric chemical depletion patterns (see Section~\ref{sec:int}) also hint at sustained gas-dust fractionation, which requires the presence of pressure maxima or other trapping mechanisms to prevent rapid radial drift of dust grains \citep{vandermarel2013DustTrapTransitionDisc}.

In this context, the azimuthal asymmetry detected in IRAS\,08544$-$4431 (structure A) provides the first spatially resolved (at physical scales close to $\sim$15\,au) observational evidence for the overdensities of mm/sub-mm-sized dust grains in face-on CBDs of post-AGB binaries. Assessing the prevalence and characteristic amplitudes of azimuthal asymmetries across the post-AGB binary population is therefore important for determining whether localized pressure structures are a generic outcome of disk-binary interaction in evolved systems, or instead require more specific dynamical conditions.

\subsection{Similarities between the CBD of IRAS 08544--4431 and Planet-Forming Disks around Young Stars}\label{ssec:dscppd}
High-resolution ALMA surveys have shown that PPDs around young stars, resolved at $<100$\,au, commonly exhibit axisymmetric rings and gaps arranged in multiple concentric patterns \citep{andrews2018DSHARP, huang2018DSHARPDiskSubstructures, long2018PPDGapsRings, yamaguchi2024DiskSubstructures}. These substructures are observed in disks around stars spanning a wide range of stellar masses and metallicities, indicating that the formation of localized dust structures is a common outcome of PPD evolution \citep{vandermarel2019PPDStructuresAcrossAgesLuminosities, huang2024PPDsInSigmaOri, vioque2025PPDSubstructuresAGEPRO}. Furthermore, surveys with ALMA, which trace midplane emission, and SPHERE/VLT, which probes scattered light from disk surfaces, revealed that PPDs with spiral arms and azimuthal asymmetries, including crescent-like overdensities, constitute a significant minority of systems \citep[observed in 10-25\% of PPDs;][]{huang2018DSHARPDiskSubstructures, dong2018SpiralArmsPPDsSPHERE}.

In PPDs, planet-disk interactions are thought to generate spiral arms and azimuthal asymmetries by carving gaps, producing dust-trapping pressure maxima, inducing eccentricity in the disk, and creating asymmetric overdensities or vortices \citep{ataiee2013PlanetInducedDiskEccentricity, vandermarel2013DustTrapTransitionDisc, long2018PPDGapsRings, huang2018DSHARPDiskSubstructures, cieza2021ODISEA, fedele2021PlanetFormationinPPD}. Alternative mechanisms capable of producing rings and azimuthal structure in the inner parts of the disk include variations in dust growth and fragmentation, photo-evaporation, and dead zones due to magneto-rotational instabilities \citep{flock2015PPDSimulations, dullemond2018DustTrapping, wu2023MRI, yamaguchi2024DiskSubstructures}. The diversity of morphologies and radii at which these mechanisms operate \citep[20-100\,au;][]{vandermarel2013DustTrapTransitionDisc} suggests that several processes may act in parallel within a PPD. Importantly, observational studies indicate that PPD substructures emerge early \citep[$\sim 3\times10^5$ years;][]{hsieh2025SubstructureEmergencePPDs} and persist throughout the typical disk lifetime \citep[$10^6-10^7$\,years; e.g.,][]{huang2018DSHARPDiskSubstructures, yamaguchi2024DiskSubstructures, vioque2025PPDSubstructuresAGEPRO}. Narrow, optically thick rings provide favorable conditions for dust trapping \citep{dullemond2018DustTrapping} and planetesimal formation \citep[see][and references therein]{andrews2018DSHARP}, while azimuthal asymmetries are often interpreted as vortices or dust concentrations, frequently associated with planet-induced perturbations \citep{flock2015PPDSimulations, yang2023PPDVortices, wu2023MRI}. The existence of these asymmetric features in a significant minority of PPDs highlights the efficiency of processes that generate localized dust overdensities in PPDs.

The analogy between PPDs of young stars and CBDs of post-AGB binaries might be not only morphological, but also causal. Although CBDs of post-AGB binaries are primarily governed by disk–binary interactions rather than planet-formation processes \citep{vanwinckel2003PostAGBReview, vanwinckel2025Review}, these CBDs and PPDs \citep[including circumbinary PPDs; see, e.g.,][and references therein]{vioque2026CircumbinaryPPDs} share key ingredients, including the fractionation of gas and dust, substantial grain growth, inner dust cavities, and aspect ratios of $\frac{H}{r}\sim0.18$ \citep{flock2015PPDSimulations, andrych2023DiscStructuresIRDIS, vanwinckel2025Review, vioque2025PPDSubstructuresAGEPRO, byrne2026AspectRatiosPPDs}. This analogy has led to CBDs of post-AGB binaries being commonly referred to as second-generation PPDs. For IRAS\,08544$-$4431, the compact overdensity at $r\approx45$\,mas (structure A), together with weaker asymmetries (see Section~\ref{sec:res}), resembles crescent-like structures observed in PPDs \citep[including those with ages below 10$^6$ years;][]{vandermarel2021DiskAsymmetriesPPDs}, and is consistent with a localized pressure maximum capable of slowing radial drift and, potentially, azimuthal drift of dust grains relative to the gas phase. Furthermore, the conservative derived range of gas-to-dust ratios for the CBD of IRAS\,08544$-$4431 ($\sim$10-200; see Section~\ref{ssec:resflx}) might favor the second-generation planet formation within this CBD \citep[e.g., via streaming instability;][]{pourmand2025GravInstability}.

The detection of dust asymmetries in this study (structures A, B, C, and D) reinforces that analogous dust-concentration processes should operate in disks across a broad range of stellar evolutionary stages, from long-lived ($\gtrsim10^6$ years) PPDs around young stars to short-lived ($10^4$-$10^5$ years) CBDs around evolved binaries. In this context, the extensive multi-wavelength observations of IRAS\,08544$-$4431 obtained with SPHERE/VLT \citep{andrych2024IRAS08}, PIONIER/VLT \citep[][De Prins et al., in prep.]{hillen2016IRAS08}, MATISSE/VLT \citep{corporaal2023FullDisc}, GRAVITY/VLT (Corporaal et al., in prep.), and ALMA (this study) provide a uniquely rich empirical framework for investigating dust-trapping processes in CBDs around post-AGB binaries. In future work, we will combine this multi-technique observational framework with \texttt{MCFOST} radiative-transfer modelling \citep{pinte2009MCFOST} and \texttt{PHANTOM} hydrodynamical simulations \citep{price2018PHANTOM} to quantitatively explore the physical mechanisms responsible for the observed asymmetries.

\section{Conclusions}\label{sec:con}
In this study, we analyzed the highest-angular-resolution ALMA Band 7 continuum and $^{12}$CO $J$ = 3-2 observations obtained to date for IRAS\,08544$-$4431, resolving the CBD of this system at $\sim$10\,mas. These data probe the inner disk at spatial scales close to the reported sublimation rim diameter ($\sim$14\,mas) and reveal substructures that remained inaccessible in previous observations. The ALMA 2021 continuum map in 0.87-mm regime shows that the dust distribution in the disk midplane of IRAS\,08544--4431 is mostly smooth with prominent non-axisymmetric features (peaking at 20-25\% of the flux density in the smooth profile at the corresponding radii, which corresponds to $>$6$\times$RMS). We detect a compact azimuthal overdensity at $r\approx45$\,mas, with an azimuthal extent of $\sim$50\,mas, accompanied by a relative depression in the flux density on the opposite (southeastern) side of the disk. The location of this feature is consistent with a localized pressure perturbation capable of concentrating dust grains with mm-/sub-mm sizes. Additional weaker residuals at other radii further suggest that the disk hosts multiple perturbations in addition to mostly smooth radial structure.

The $^{12}$CO line map from the 2021 observation corroborates with previously established global Keplerian rotation of the disk, while the $^{12}$CO moment 0 map hints that the CO distribution might not spatially coincide with the continuum overdensity. This indication of an offset further suggests that gas and dust are not strictly co-spatial but rather dynamically decoupled in the disk midplane. This behavior is qualitatively consistent with dust trapping in a pressure maximum, where the local gas-to-dust ratio is decreased relative to the background disk. We note, however, that higher-sensitivity ALMA data are required to establish the offset robustly.

Considered together with recent high-resolution ALMA observations of the Red Rectangle and the commonly observed photospheric depletion in post-AGB binaries, our results suggest that azimuthal asymmetries might be omnipresent within the post-AGB binary population, closely similar to the substructures commonly observed in protoplanetary disks. Our findings demonstrate the significant efficiency of pressure-mediated dust concentration in circumbinary environments of post-AGB binaries shaped by disk-binary interaction. Confirming whether such asymmetries of dust distribution are ubiquitous, and determining their characteristic amplitudes and lifetimes, will be important for assessing whether dust trapping is a generic outcome of rotationally supported disks across evolutionary stages, or whether it arises only under specific dynamical conditions. Future ALMA observations of dust continuum and gas lines will be crucial for testing how typical the detected overdensities in the CBDs of IRAS\,08544$-$4431 and the Red Rectangle are for the post-AGB binary population, thereby constraining the mechanisms of disk-binary interaction in these systems.

%% Please use the acknowledgment and contribution environments. This will 
%% be anonymized when the "anonymous" style option is used. 
\begin{acknowledgments}
This paper makes use of ALMA observations with the following IDs: 2019.1.00919.S and 2013.1.00338.S. ALMA is a partnership of ESO (representing its member states), NSF (USA) and NINS (Japan), together with NRC (Canada), NSTC and ASIAA (Taiwan), and KASI (Republic of Korea), in cooperation with the Republic of Chile. The Joint ALMA Observatory is operated by ESO, AUI/NRAO and NAOJ.

MM, DK, HVW, and KA acknowledge the support of the Australian Research Council discovery Project DP240101150. TDP acknowledges support of the Research Foundation - Flanders (FWO) under grant 11P6I24N. This project has received funding from the European Research Council (ERC) under the European Union Horizon Europe programme (grant agreement No. 101042275, project Stellar-MADE). JA and VB acknowledge partial support from project CRISPNESS, grant PID2023-146056NB-C21, funded by MICIU/AEI/10.13039/501100011033 and by ERDF/EU. NC acknowledges funding from the European Research Council (ERC) under the European Union Horizon Europe programme (grant agreement No. 101042275, project Stellar-MADE).
\end{acknowledgments}

\begin{contribution}
%%This section gives authors the space to recognize author contributions. The text inside this environment is NOT counted towards the total word quanta. At a minimum, manuscripts are expected to include this text:

MM was responsible for analyzing and interpreting ALMA data and for writing and submitting the manuscript.
DK, JK, and HVW came up with the initial research concept and edited the manuscript.
EC contributed to the development of the analysis pipeline.
KA contributed to the formal analysis and validation.
DP, TDP, ACC, JA, JK, NC, and MF contributed to the interpretation of the obtained results.
JK, ACC, JA, DK, HVW, and VB secured the ALMA 2021 dataset analyzed in this study.
%%
%% Authors can use the Contributor Role Taxonomy (CRediT) at
%% https://credit.niso.org
%% for ideas on how write a good statement tailored to their needs.

\end{contribution}

%% To help institutions obtain information on the effectiveness of their 
%% telescopes the AAS Journals has created a group of keywords for telescope 
%% facilities.
%
%% Following the acknowledgments section, use the following syntax and the
%% \facility{} or \facilities{} macros to list the keywords of facilities used 
%% in the research for the paper.  Each keyword is check against the master 
%% list during copy editing.  Individual instruments can be provided in 
%% parentheses, after the keyword, but they are not verified.
\facilities{ALMA}

%% Similar to \facility{}, there is the optional \software command to allow 
%% authors a place to specify which programs were used during the creation of 
%% the manuscript. Authors should list each code and include either a
%% citation or url to the code inside ()s when available.
\software{CASA, own codes}

%% Appendix material should be preceded with a single \appendix command.
%% There should be a \section command for each appendix. Mark appendix
%% subsections with the same markup you use in the main body of the paper.
%%
%% Each Appendix (indicated with \section) will be lettered A, B, C, etc.
%% The equation counter will reset when it encounters the \appendix
%% command and will number appendix equations (A1), (A2), etc. The
%% Figure and Table counter will not reset.

\appendix

\section{Impact of Weighting Choice on Combined ALMA Maps of IRAS 08544--4431}\label{app:rob}
In this Appendix, we provide the ALMA maps of continuum (and the respective continuum maps with azimuthally-averaged disk profile subtracted) and $^{12}$CO (3-2) line for the CBD of IRAS\,08544$-$4431 obtained from the combined (2015, 2016, and 2021) observational dataset.

\begin{figure}[htb]
    \centering
    \includegraphics[width=0.32\textwidth]{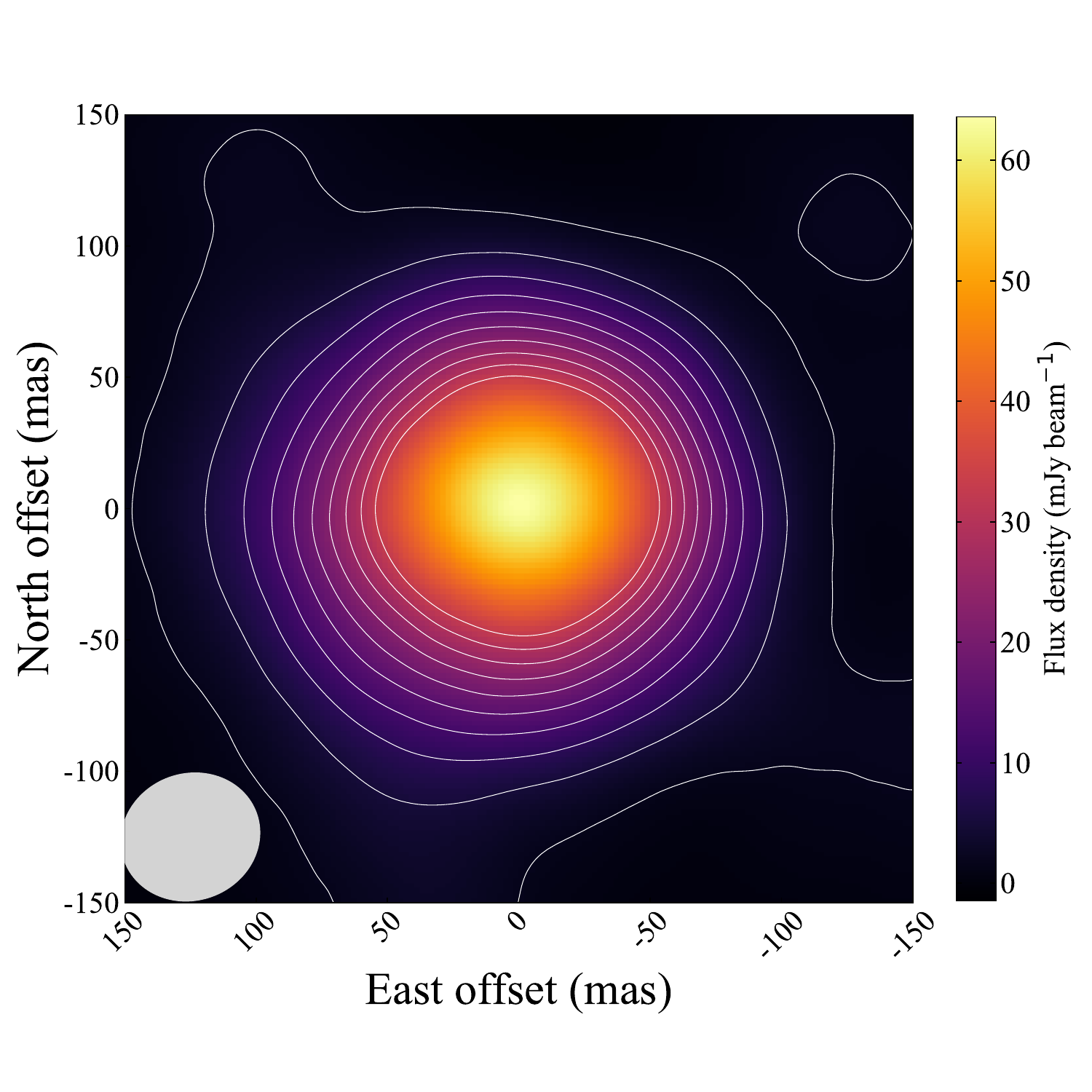}
    \includegraphics[width=0.32\textwidth]{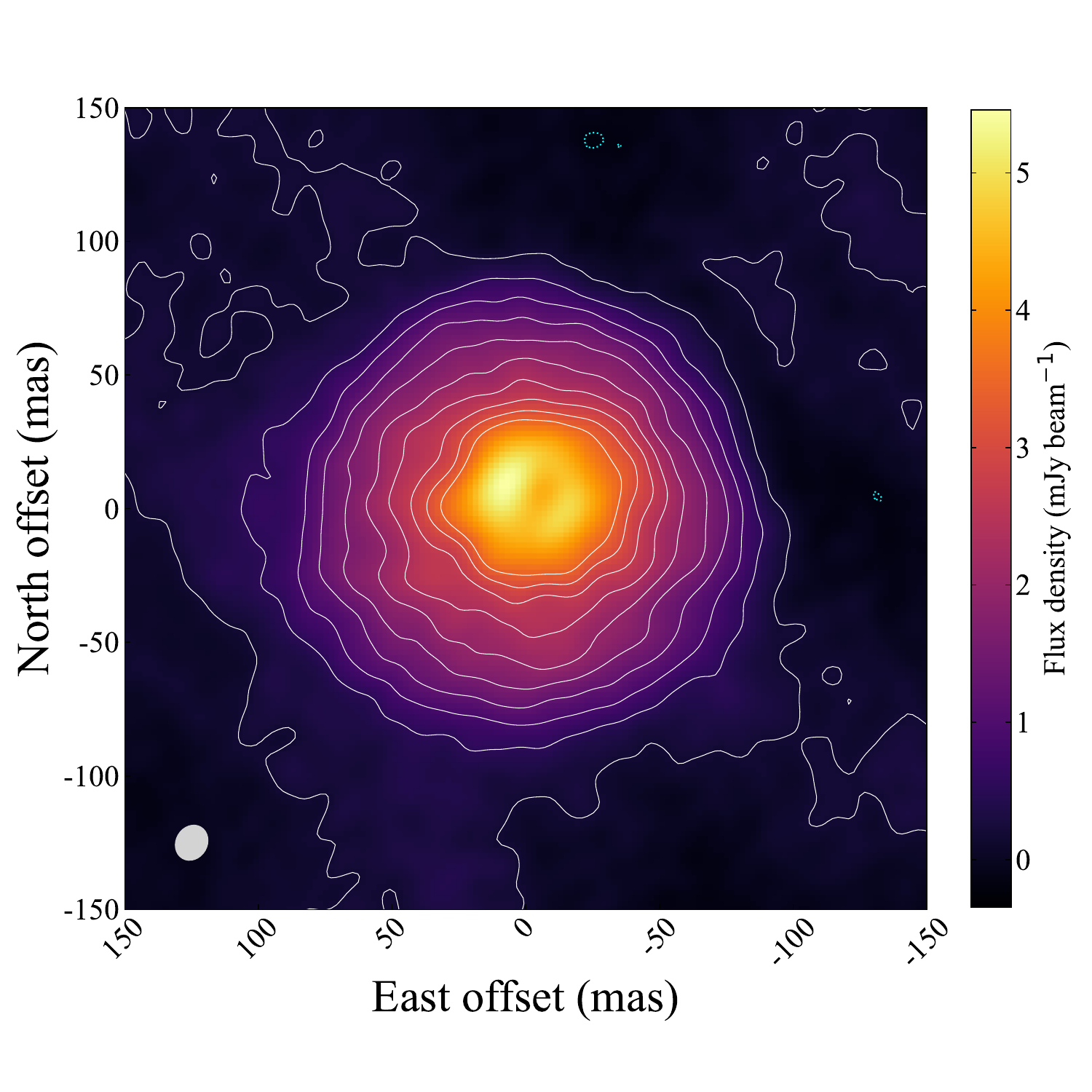}
    \includegraphics[width=0.32\textwidth]{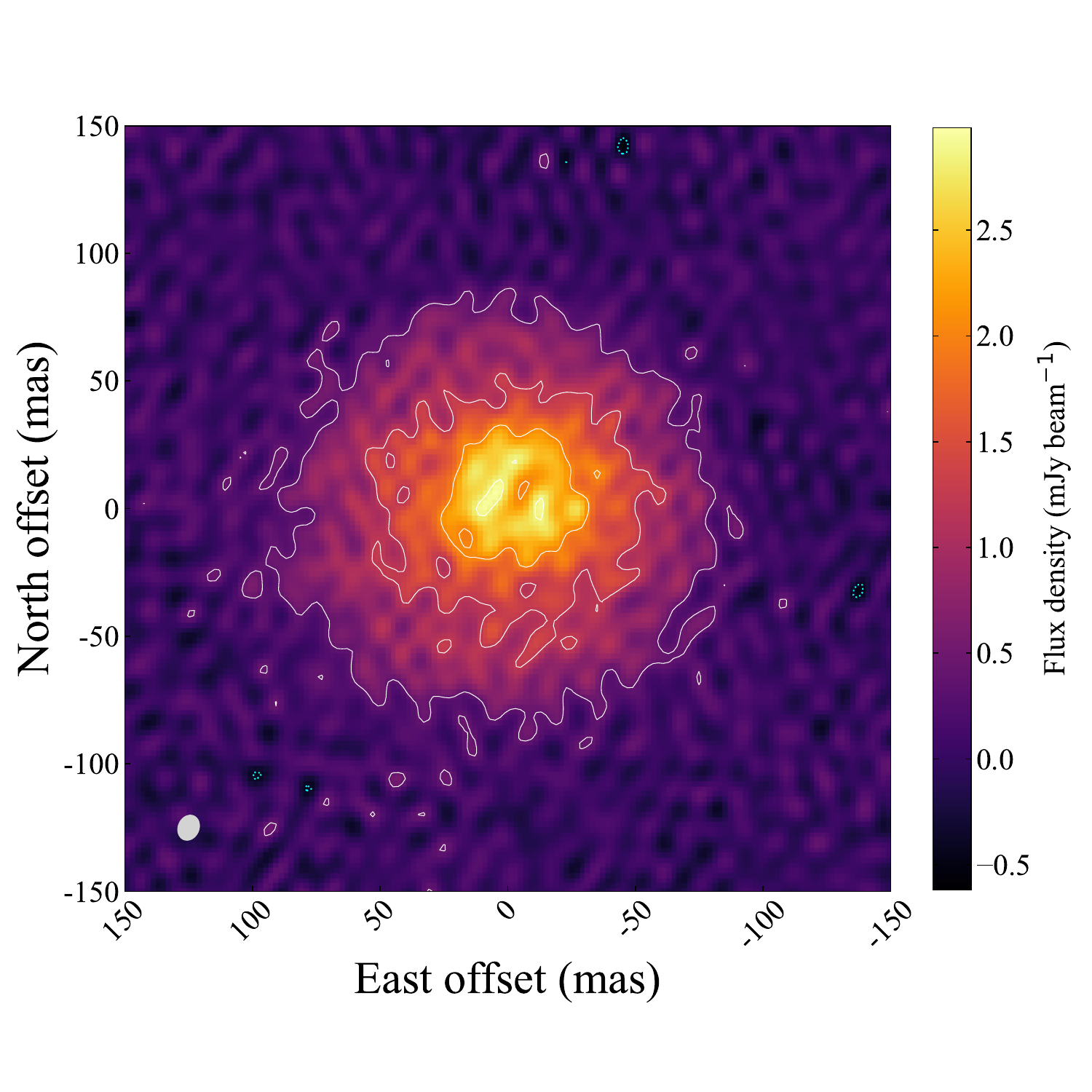}
    \includegraphics[trim={12.5cm 0 0 0}, clip, width=0.32\textwidth]{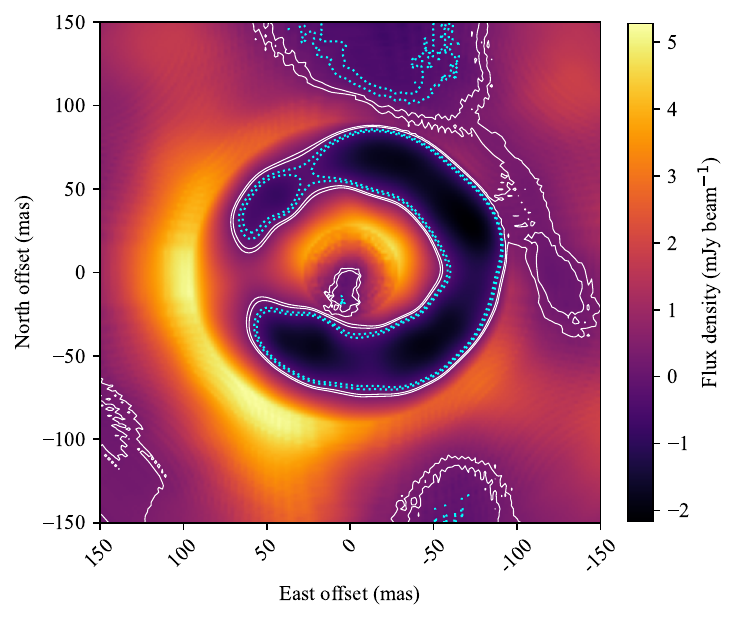}
    \includegraphics[trim={12.5cm 0 0 0}, clip, width=0.32\textwidth]{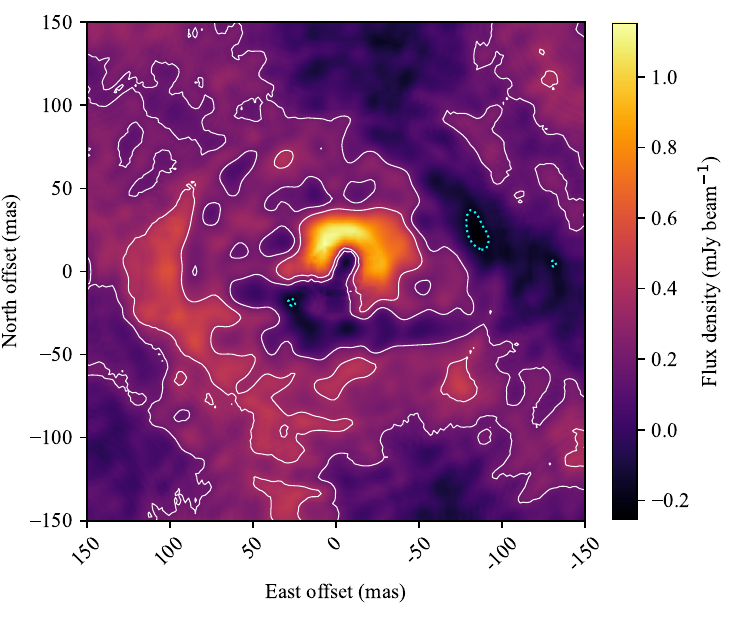}
    \includegraphics[trim={12.5cm 0 0 0}, clip, width=0.32\textwidth]{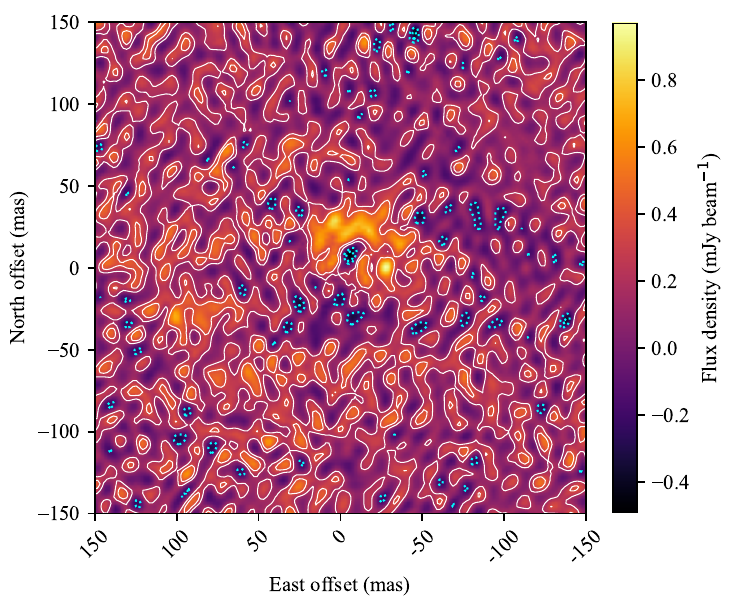}
    \caption{Continuum maps (top panels) and subtracted maps (bottom panels) of IRAS\,08544$-$4431 plotted from the combined ALMA data from 2015, 2016, and 2021 observations using three different weightings: natural (left panels), Briggs with robust=0 (middle panels), and uniform (right panels). The beam size and orientation are shown in the bottom left corner. The contours mark the similar multiples of RMS values (0.573, 0.06, and 0.136 mJy beam$^{-1}$ for natural, Briggs robust=0, and uniform weightings, respectively), as in Fig.~\ref{fig:cntmap} (for more details, see Appendix~\ref{app:rob}).}\label{fig:appwgt}
\end{figure}

For the continuum, we plotted the maps using natural, Briggs, and uniform weightings (left, middle, and right panels, respectively; see Fig.~\ref{fig:appwgt}). The axisymmetric distribution of dust emission is most prominent in Briggs and uniform weightings of the combined maps, though located in the southeastern direction. We note that this inconsistency of dust overdensity location with 2021 observation might stem from a possible shift between the center coordinates for 2015, 2016, and 2021 observations ($\sim$10-20\,mas), which will be further explored in the next paper dedicated to the radiative-transfer modelling of the CBD in IRAS\,08544$-$4431.

\begin{figure}[htb]
    \centering
    \includegraphics[trim={0.25cm 0.1cm 0.25cm 0.25cm},width=0.675\textwidth]{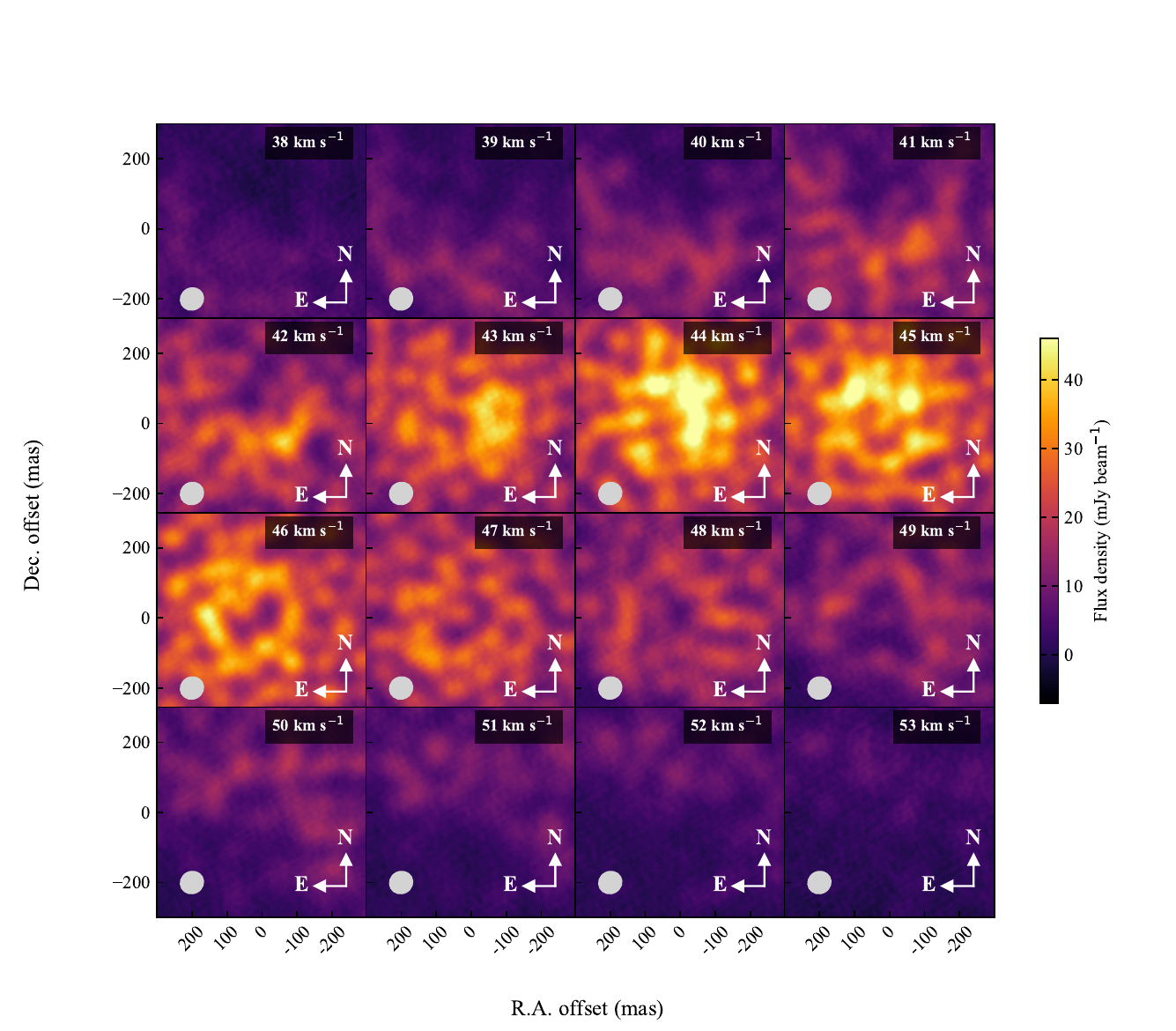}
    \includegraphics[trim={0.25cm 0.25cm 0.25cm 0.25cm},width=0.675\textwidth]{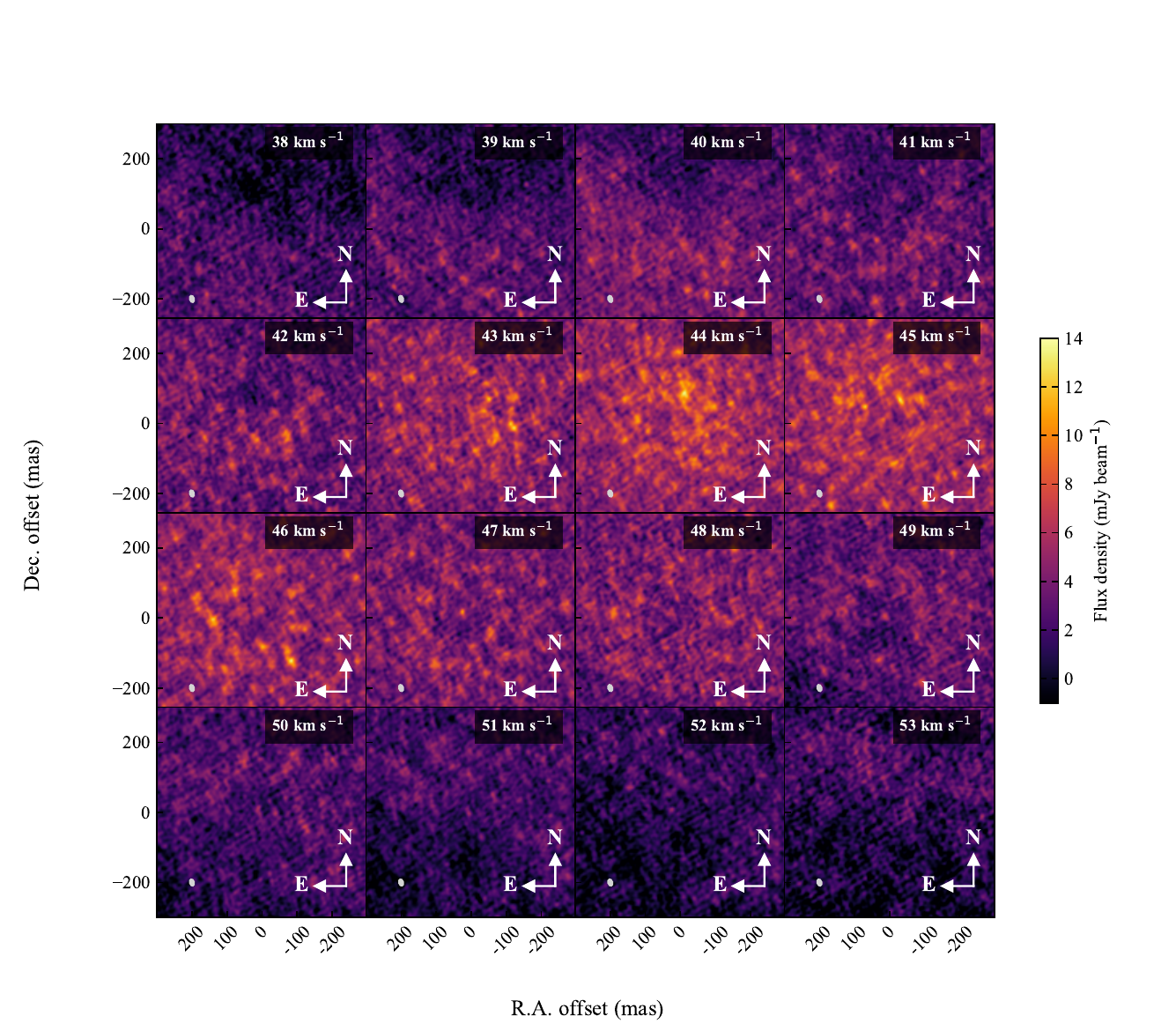}
    \caption{$^{12}$CO line maps of IRAS\,08544$-$4431 plotted from the combined ALMA data from 2015, 2016, and 2021 observations using two different weightings: natural (top) and Briggs with robust=0.5 (bottom). The beam size and orientation are shown in the bottom left corner (for more details, see Appendix~\ref{app:rob}).}\label{fig:appco1}
\end{figure}

\begin{figure}[htb]
    \centering
    \includegraphics[trim={0.25cm 0.1cm 0.25cm 0.25cm},width=0.675\textwidth]{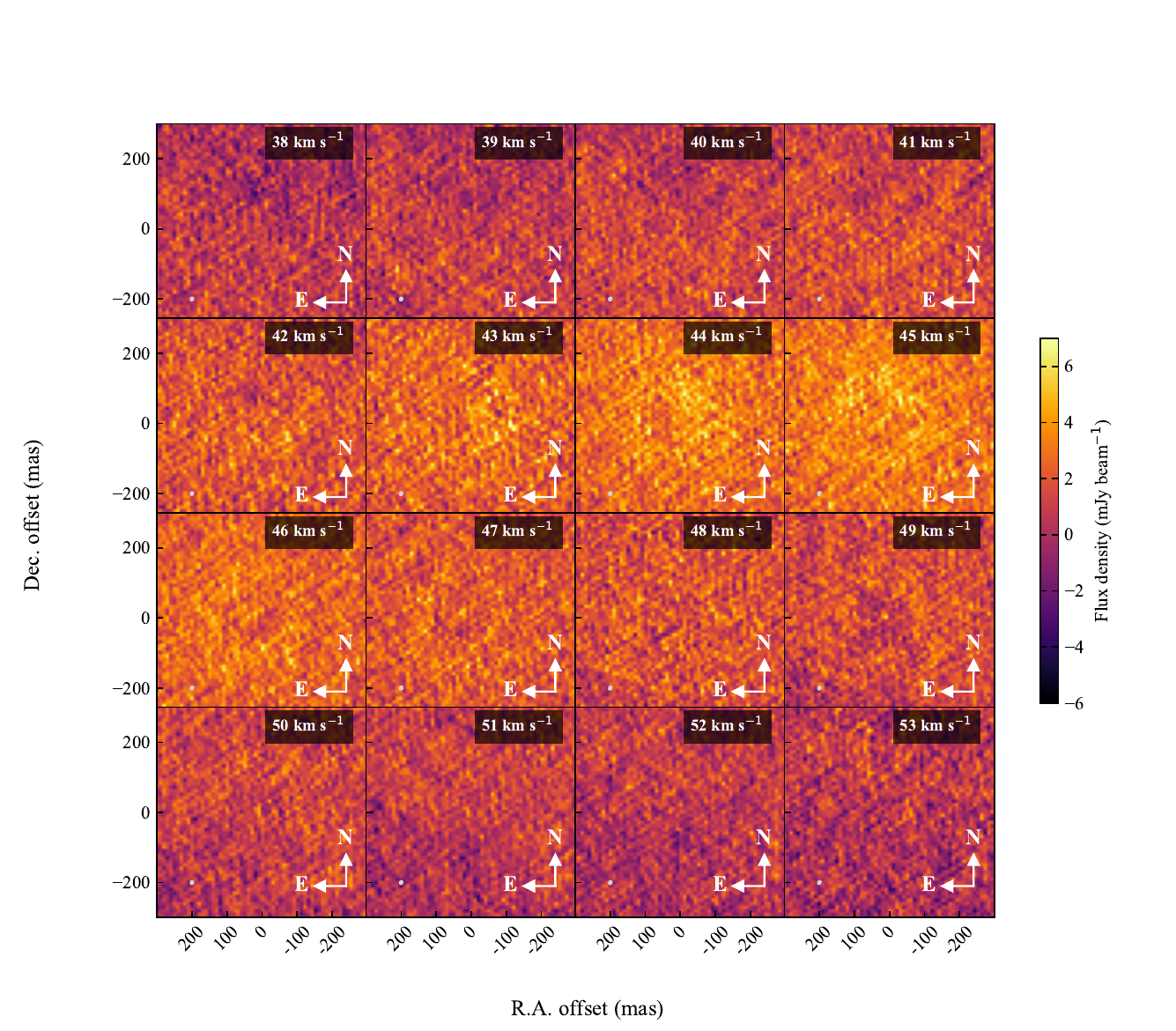}
    \includegraphics[trim={0.25cm 0.25cm 0.25cm 0.25cm},width=0.675\textwidth]{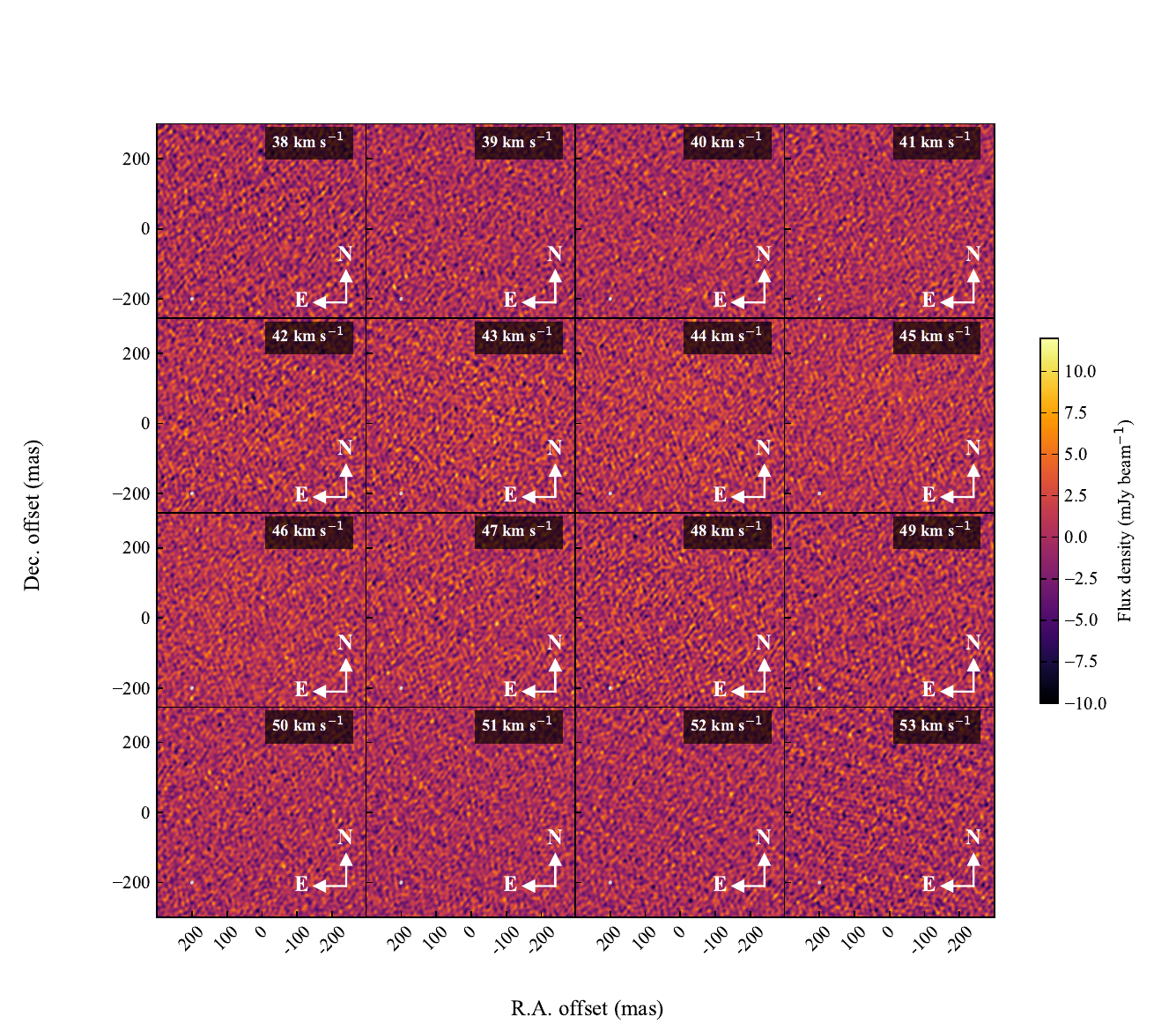}
    \caption{$^{12}$CO line maps of IRAS\,08544$-$4431 plotted from the combined ALMA data from 2015, 2016, and 2021 observations using two different weightings: Briggs with robust=0 (top) and uniform (bottom). The beam size and orientation are shown in the bottom left corner (for more details, see Appendix~\ref{app:rob}).}\label{fig:appco2}
\end{figure}

For the $^{12}$CO line, we plotted the maps using natural, Briggs r=0.5, Briggs r=0, and uniform weightings (see Fig.~\ref{fig:appco1} and \ref{fig:appco2}). We conclude that the axisymmetric features observed in ALMA 2021 observation are smoothed out in the combined dataset.

 % The field orientation is shown in the bottom right corner.

 % The field orientation is shown in the bottom right corner

 % The field orientation is shown in the bottom right corner

\section{Surface and Midplane Structure of the CBD in IRAS 08544--4431}\label{app:sph}
In this Appendix, we present the comparison of the high-angular-resolution data for the CBD in IRAS\,08544$-$4431 obtained with SPHERE/VLT \citep{andrych2024IRAS08}, PIONIER/VLTI \citep{hillen2016IRAS08, kluska2018IRAS08}, and ALMA (this study).

\begin{figure}[htb]
    \centering
    \includegraphics[width=\textwidth]{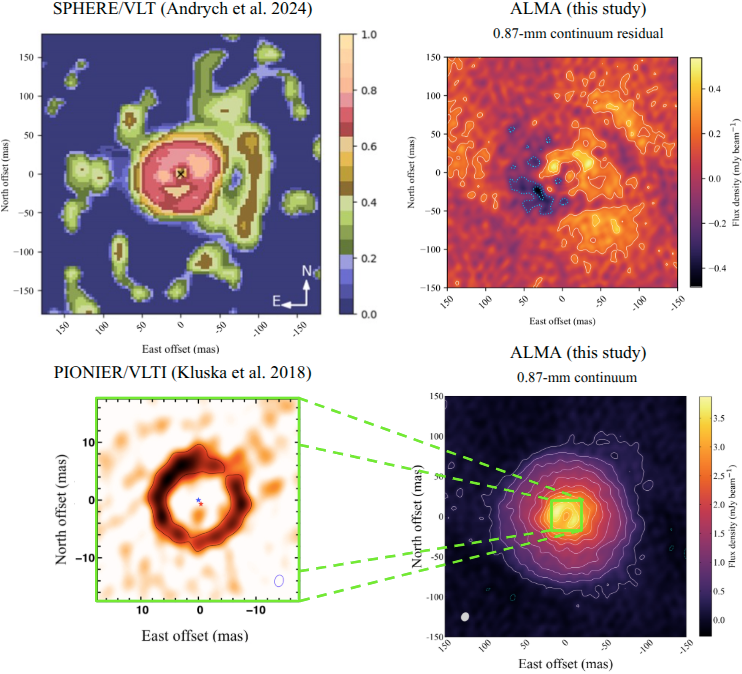}
    \caption{Comparison of disc morphologies discovered in the CBD of IRAS 08544$-$4431. \textit{Top left panel:} combined $Q_\varphi$ polarized disc morphology using V-, I-, and H-band data from SPHERE/VLT \citep[presented on an inverse hyperbolic scale and normalized to highlight the intensity change along the disc;][]{andrych2023DiscStructuresIRDIS, andrych2024IRAS08}. \textit{Top right panel:} ALMA residual map after subtracting radially symmetric model (this study). \textit{Bottom left panel}: image reconstruction of the PIONIER/VLTI dataset \citep{kluska2018IRAS08}. \textit{Bottom right panel:} ALMA continuum map at 0.87 mm (this study). For more details, see Appendix~\ref{app:sph}.}\label{fig:appplt}
\end{figure}

In the top panels of Fig.~\ref{fig:appplt}, we compare $Q_\varphi$ polarized image of the CBD in IRAS\,08544$-$4431 \citep[top left panel;][observed on 25 October 2018]{andrych2024IRAS08} with our ALMA continuum residuals (top right panel; observed on 5 September 2021). SPHERE/VLT dataset of IRAS\,08544$-$4431 revealed a pronounced forward scattering peak in the NW direction \citep[see Fig.~12 in][observed on 25 October 2018]{andrych2024IRAS08}, while ALMA 2021 dataset reveals a clear asymmetry in the disk midplane located in the same azimuth range, though the center of the dominant overdensity appears to partially overlap with a small ($\sim$20$\times$10\,mas) local minimum of $Q_\varphi$ polarized intensity. Although the spatial correspondence is striking, the two features likely originate from different physical processes. The forward scattering peak reflects the geometry and structure of the disk surface, showing the illuminated front side of the CBD, whereas ALMA 0.87-mm continuum traces density variations close to the midplane that are expected to orbit the central binary. The alignment should therefore be coincidental rather than evidence of a direct link between the surface scattering and the (almost) midplane dust distribution.

In the bottom panels of Fig.~\ref{fig:appplt}, we extend our multi-technique comparison of the CBD in IRAS\,08544$-$4431 by linking the PIONIER/VLTI data tracing the disk inner rim \citep[bottom left panel;][]{kluska2018IRAS08} with the ALMA 2021 continuum map presented in this study (bottom right panel). The characteristic spatial scale of the inner rim emission detected by PIONIER/VLT is consistent with the inner boundary of the ALMA continuum distribution, indicating that the observed 0.87-mm emission arises from regions immediately adjacent to the dust sublimation rim. This spatial correspondence supports the interpretation that the ALMA 2021 observations probe the disk midplane down to radii comparable to the inner rim, thereby providing a direct link between the structure of the innermost disk layers at the sublimation rim and at further radii.

%Combined polarized disc morphology for IRAS 08544–4431 using V- and I-band data from this study, along with the SPHERE/IRDIS H-band results (adapted from Andrych et al. 2023). The black cross represents the position of the binary star. Before combining the images in each band, they were normalized to the total intensity, corrected for the separation-dependent drop-off in illumination by multiplying by and accounting for the system inclination defined for the resolved disc surface in each band (see Table 2). Additionally, the image is presented on an inverse hyperbolic scale and normalized to highlight the intensity change along the disc. See Section 4.4 for more details.

\bibliography{master}{}
\bibliographystyle{aasjournalv7}

%% This command is needed to show the entire author+affiliation list when
%% the collaboration and author truncation commands are used.  It has to
%% go at the end of the manuscript.
%\allauthors

%% Include this line if you are using the \added, \replaced, \deleted
%% commands to see a summary list of all changes at the end of the article.
%\listofchanges

\end{document}